\documentclass[a4paper,12pt]{article}

\usepackage[english]{babel}
\usepackage{amsmath,amssymb,amsbsy,amstext}
\usepackage[dvips]{graphicx}
\usepackage[nospace]{cite}
\usepackage{amssymb}
\usepackage{mathrsfs}
\usepackage{psfrag}
\usepackage{subfigure}
\usepackage[sc,small]{caption2}
\setcaptionwidth{14cm}
 
\newcommand{\be}{\begin{eqnarray}}
\newcommand{\non}{\nonumber \\}
\newcommand{\V}{\mathcal{V}}
\newcommand{\CO}{\mathcal{O}}
\newcommand{\expo}{\beta}
\newcommand{\gk}{g_{\rm K}}
\newcommand{\gw}{g_{\rm W}}

\newcommand{\ee}{\end{eqnarray}}
\newcommand{\bdm}{\begin{displaymath}}
\newcommand{\edm}{\end{displaymath}}
   
\newcommand{\ba}{\begin{array}}
\newcommand{\ea}{\end{array}}

\newcommand{\E}{\mathcal{E}}    
\newcommand{\Et}{\mathcal{E}^{(K)}}
\newcommand{\Es}{\mathcal{E}^{(W)}}
\newcommand{\al}{\alpha'}

\newcommand{\W}{W_{\mathrm{np}}}
\newcommand{\Ub}{\bar{U}}

\newcommand{\ca}{{\cal A}}
\newcommand{\ce}{{\cal E}}
\newcommand{\ck}{{\cal K}}

\newcommand{\cm}{{\cal M}}

\newcommand{\cf}{{\cal F}}

\newcommand{\cv}{{\cal V}}

\newcommand{\co}{{\cal O}}

\newcommand{\cq}{{\cal Q}}

\newcommand{\cz}{{\cal Z}}
\newcommand{\tr}{{\rm tr}}

\newcommand{\zba}[2]{[\!\!\begin{array}{c}{\scriptstyle#1}%
                        \\[-1.6mm]{\scriptstyle #2}\end{array}\!\!]}

\newcommand{\thbw}[2]{\vartheta[\genfrac{}{}{0pt}{}{#1}{#2} ]}

\begin{document}

\title{}
\author{}
\date{}
\thispagestyle{empty}

\begin{flushright}
\vspace{-3cm}
{\small LMU-ASC-21/07 \\
        AEI-2007-020}
\end{flushright}
\vspace{1cm}

\begin{center}
{\bf\LARGE
Jumping Through Loops:\\[2mm]
On Soft Terms from Large Volume Compactifications}

\vspace{1.5cm}

{\bf Marcus Berg}$^{+}$
{\bf,\hspace{.2cm} Michael Haack}$^{\dag}$
{\bf \hspace{.1cm} and \hspace{.2cm} Enrico Pajer}$^{\dag}$
\vspace{1cm}

{\it
$^{+}$ Max Planck Institute for Gravitational Physics\\ %
Albert Einstein Institute\\  %
M\"uhlenberg 1, D-14476 Potsdam, Germany \\[24pt] %
$^{\dag}$ Ludwig-Maximilians-University \\ %
Department of Physics \\ %
Theresienstr. 37, D-80333 M\"unchen, Germany 
}

\vspace{1cm}

{\bf Abstract}
\end{center}
\vspace{.5cm}
We subject
the phenomenologically successful 
large volume scenario of hep-th/0502058 to 
a first consistency check in string theory. 
In particular, we consider whether the 
expansion of the
string effective action is 
consistent in the presence of D-branes and O-planes.  
Due to the no-scale structure at tree-level, the 
scenario is surprisingly robust.  
We compute the
modification of soft supersymmetry breaking terms,
and find only subleading corrections. We also comment that
for large-volume limits of toroidal 
orientifolds and fibered Calabi-Yau manifolds
the corrections can be more important,
and we discuss further checks that need to be performed.
\clearpage


\tableofcontents


\section{Introduction}

The KKLT strategy \cite{Giddings:2001yu,Kachru:2003aw}
for producing stabilized string vacua that can serve as a
starting point for phenomenology has been a source of great interest
for the last few years.
The ``large-volume scenario'' (LVS)
\cite{Balasubramanian:2005zx,Conlon:2005ki}
is an extension of KKLT where string
corrections to the tree-level supergravity effective action 
computed in \cite{Becker:2002nn} play a
significant role, and where the compactification volume 
can be as large as $10^{15}$ in string units.
In LVS, work has been done on 
soft supersymmetry breaking 
\cite{Conlon:2005ki,Conlon:2006us,Conlon:2006wz}, the QCD axion
\cite{Conlon:2006tq,Conlon:2006ur}, neutrino masses
\cite{Conlon:2006wt}, inflation 
\cite{Conlon:2005jm,Holman:2006ek,Simon:2006du,Bond:2006nc}, 
and even first attempts at
LHC phenomenology \cite{Kane:2006yi}. 

Although tantalizing, the models
discussed in the aforementioned papers 
(nominally ``string compactifications'')
raise many questions. 
It remains an open problem to construct complete KKLT models
in string theory, as 
opposed to supergravity. 
Problems one faces include things like
the description of RR fluxes in string theory,
showing that the necessary nonperturbative effects actually can and do
appear in a way consistent with other contributions to the potential
(for progress in
this direction, see \cite{Denef:2004dm,Gorlich:2004qm,Tripathy:2005hv,Denef:2005mm,Saulina:2005ve,Kallosh:2005gs,Martucci:2005rb,Berglund:2005dm,Bergshoeff:2005yp,Lust:2005cu,Lust:2006zg,Blumenhagen:2006xt,Haack:2006cy,Akerblom:2006hx,Tsimpis:2007sx,Bianchi:2007fx,Argurio:2007vq}), 
and verifying that one can uplift to a Minkowski
or deSitter vacuum without ruining stabilization 
\cite{Choi:2004sx,Lust:2005dy,Krefl:2006vu,Lust:2006zg,Gomez-Reino:2006dk}. 
In LVS, since string corrections play a
crucial role, striving for actual string constructions seems
quite important. In the end, the restrictiveness this entails may
greatly improve predictivity, or kill the models completely 
as string compactifications.

In this paper, we will not improve on
the consistency of KKLT or LVS in general, but rather assume 
the existence of LVS models
in string theory, and then perform self-consistency checks.
This is a modest step on the way towards reconciling phenomenologically
promising scenarios with underlying string models.
We will see that although a priori the situation looks very
bleak, and one might have hastily concluded that even our modest
consistency check would put very strong constraints on LVS, 
things are more interesting. It turns out that LVS jumps through every hoop 
we present it with, and instead of broad qualitative changes, we find only small quantitative
changes.

The main difference between KKLT and LVS 
is that LVS includes a specific string $\alpha'$ correction
$\Delta K_{\alpha'}$ in the K\"ahler potential $K$
of the four-dimensional ${\cal N}=1$ effective supergravity. 
Naturally, the four-dimensional string effective action also contains other
string corrections. Here, we will focus on $g_{\rm s}$ corrections
due to sources (D-branes and O-planes). For some ${\cal N}=1$ and ${\cal N}=2$ toroidal orientifolds, 
these corrections were computed in \cite{Berg:2005ja} (see also 
\cite{Antoniadis:1996vw}; for a comprehensive introduction
to orientifolds, see \cite{Angelantonj:2002ct}). 
Compared to the $\alpha'$ correction $\Delta K_{\alpha'} $ considered in LVS, the $g_{\rm s}$ corrections 
to the K\"ahler potential $ \Delta K_{g_{\rm s}}$ will
scale as
%
%
%
\be   \label{firstscaling}
\Big(\Delta K_{\alpha'} : \Delta K_{g_{\rm s}}\Big) \sim \Big(\CO(\alpha'^3) : 
\CO(g_{\rm s}^2 \alpha'^2)\Big)
\qquad \mbox{(string frame)}\ .
\ee
By naive dimensional analysis, one would expect that
in a $1/\V$ expansion, where $\V$ is the overall volume 
in the Einstein frame,  eq.\ \eqref{firstscaling} implies
%
%
\be \label{orders2}
\Delta K_{\alpha'} \;\sim\; \CO(g_{\rm s}^{-3/2}\V^{-1})\quad , 
\quad \Delta K_{g_{\rm s}} \;\sim\; \CO(g_{\rm s} \V^{-2/3})\
\qquad \mbox{(Einstein frame)}\ .
\ee
If there is more than one K\"ahler modulus, as is usually the case, 
various combinations of K\"ahler moduli may 
appear in $\Delta K_{g_{\rm s}}$ in eq.\eqref{orders2}, and a priori this could lead to
even {\it weaker} suppression in
$1/\V$ than that shown. However, we will argue that \eqref{orders2} is actually correct as far as the
suppression factors in the $1/\V$ expansion go. Nevertheless, even the suppression displayed in (\ref{orders2}) 
seems to be a challenge for LVS, if indeed $\V \sim 10^{15}$. For $\V$ this large, $\Delta K_{g_{\rm s}}$ would dominate
 $\Delta K_{\alpha'}$, since we do not 
expect the string coupling $g_{\rm s}$ to be stabilized
extremely small.
On the other hand, if we are interested in the effects $g_{\rm s}$
corrections may have on the existence of the large volume minima, 
the relevant quantity to look at is the {\it scalar potential} 
$V$, rather than the K\"ahler potential $K$. It turns out that certain 
cancellations in the expression for the scalar potential leave us with leading correction terms to $V$ that scale as
\be \label{orders3}
\Delta V_{\alpha'} \;\sim\; \CO(g_{\rm s}^{-1/2}\V^{-3})\quad , 
\quad \Delta V_{g_{\rm s}}  \;\sim\; \CO(g_{\rm s} \V^{-3})\ .
\ee
This is already much better news for LVS.
However, restoring numerical factors in
\eqref{orders3}, and with $g_{\rm s}$
 typically not stabilized extremely small, 
it would seem that $\Delta K_{g_{\rm s}}$ could still have a significant effect
both on stabilization and on the resultant phenomenology (like 
soft supersymmetry breaking terms, which also 
depend on the K\"ahler potential). We will see
that although this is indeed so in principle, in practice
the models we consider are surprisingly robust against
the inclusion of $\Delta K_{g_{\rm s}}$. The clearest example of this is the
calculation of gaugino masses in sec. \ref{sec:gaugino}. The result is
that for the ``11169 model'' (analyzed in \cite{Conlon:2006us}), 
the correction to the gaugino masses 
due to $\Delta K_{g_{\rm s}}$ is negligible.
Thus, for the most part, LVS survives our onslaught unscathed.

We consider this a sign that scenarios such as LVS deserve to be taken
seriously as {\it goals} to be studied in detail in string theory, 
even as the caveats above (that apply to any KKLT-like setup)
serve to remind us that
there is much work left to be done to
really understand phenomenologically viable
stabilized flux compactifications in string theory.


\section{Review}

Let us begin by a quick review of the 
KKLT and large volume scenarios. For reasons that will become clear, 
we will want to allow for more than a single 
K\"ahler modulus. 

\subsection{KKLT}

The KKLT setup  \cite{Giddings:2001yu,Kachru:2003aw} is a 
warped Type IIB flux compactification on a Calabi-Yau 
(or more generally, F-theory) orientifold,
with all moduli stabilized. In this paper, we will neglect
warping. For progress towards taking warping into account
in phenomenological contexts, see
\cite{Burgess:2006mn,Giddings:2005ff}.  

In the four-dimensional ${\cal N}=1$ effective supergravity, 
the K\"ahler potential and superpotential read
\be
K&=&-\mathrm{ln}(S+\bar{S})-2\,\mathrm{ln}(\V)+K_{\rm cs}(U,\Ub)\ , \nonumber \\
W &=& W_{\rm tree} + W_{\rm np} = W(S,U) + \sum_i A_i(S,U) e^{-a_i T_i}\ ,
\ee
where the volume $\V$ is a function of the K\"ahler moduli $T_i=\tau_i+ib_i$ whose 
real parts are 4-cycle volumes  
and whose imaginary parts are axions $b_i$,
arising from the integral of the RR 4-form over the corresponding 4-cycles. In particular, the 
volume $\V$ depends on the $T_i$ only through 
the real parts $\tau_i$, 
\be
\V=\V(T_i+\bar{T}_i) = \V(\tau_i)\ ,
\ee
and the nonperturbative superpotential $W_{\rm np}$ a priori depends
on 
the complexified dilaton
$S$ and the complex structure moduli $U$. 
After stabilization of $S$ and $U$ by 
demanding $D_U W = 0 = D_S W$, we have 
\be\label{KKLT}
K&=&-\mathrm{ln} (S+\bar{S}) -2\,\mathrm{ln}(\V)+K_{\rm cs}(U,\Ub)\ , \nonumber \\
W &=& W_{\rm tree} + W_{\rm np} = W_0 + \sum_i A_i e^{-a_i T_i}\ .
\ee
We keep the dependence on the complexified dilaton 
$S$ and the complex structure moduli $U$ in the K\"ahler potential for now, 
since the K\"ahler metric in the F-term potential 
\be \label{Fterm}
V=e^{K}\left( G^{\bar{J} I} D_{\bar{J}}\bar{W} D_I W-3|W|^2\right) 
\ee
is to be calculated 
with the full K\"ahler potential $K$, including the dependence on $S$ and $U$. (In 
eq.\ (\ref{Fterm}), the index $I$ a priori runs over all 
moduli, but after fixing the complex structure moduli and the dilaton, only the sum over
the K\"ahler moduli remains). 
The scalar potential $V$ 
has a supersymmetric AdS minimum  
at a radius that is barely large 
enough to make the use of a large-radius effective supergravity self-consistent, 
typically $\tau \sim 100$ (recall that $\tau$ has units of 
(length)${}^4$).\footnote{This minimum then has to be uplifted 
to dS or Minkowski by an additional 
contribution to the potential. Various mechanisms were suggested in 
\cite{Kachru:2003aw,Burgess:2003ic,Lebedev:2006qq,Dudas:2006vc,Dudas:2006gr,Gomez-Reino:2006dk}.}
In addition, to obtain a supersymmetric 
minimum at all, one needs to tune the flux superpotential $W_0$ to very small values.
 That is, the stabilization only works for a small parameter range. 
This is easy to understand, since we are balancing a nonperturbative term 
against a tree-level term. Let us briefly digress on
 the reasons for and implications of this balancing.


\subsection{Consistency of KKLT}
In the previous section we only considered the lowest-order supergravity effective action.
As was already noted in the original KKLT paper, $\alpha'$ corrections and 
$g_{\rm s}$ corrections (string loops) that appear in addition 
to the tree-level effective action 
could in principle affect stabilization. Oftentimes, the logic of string effective actions is 
that if one such correction matters, they all do, so no reliable physics can be learned 
from considering the first few corrections. If this is true, one can only consider 
regimes in which all corrections are suppressed.
This is not necessarily so if some symmetry prevents the tree-level contribution to 
the effective action from appearing, so that the first correction (be it $\alpha'$ 
or $g_s$) constitutes lowest order. This indeed happens for type IIB flux compactifications;
given the tree level K\"ahler potential (\ref{KKLT}), if 
we were to set $W_{\rm np}=0$, 
the remaining $K$ and $W$ in (\ref{KKLT}) produce  
a no-scale potential,
i.e.\ the scalar potential for the K\"ahler moduli then 
vanishes \cite{Cremmer:1983bf}. 
In KKLT, this no-scale
structure is only broken by the nonperturbative 
contribution to the superpotential $W_{\rm np}$. 
Since each term in $W_{\rm np}$ is exponentially suppressed in some K\"ahler modulus, the 
resulting terms in the potential are also exponentially suppressed. 
For instance, for the simpler example of a single modulus 
$\tau$, the potential (after already fixing the axionic partner along the lines of 
appendix \ref{sec:many}) reads
\be
\frac{V} {e^K}= \left[ 4|A|^2 a \tau  e^{-a\tau}\left(\frac{1}{3}a\tau+1  \right)-4 a\tau |A W_0| \right] e^{-a\tau}\ ,
\ee
meaning that even for moderate values of the K\"ahler modulus $\tau$,
all these terms are numerically very small. Corrections in $\alpha'$
and $g_s$, however, are expected to go as powers of K\"ahler moduli $\tau$, so will {\it dominate} the scalar potential for most of parameter space. In particular, it was argued in 
\cite{Balasubramanian:2005zx,Conlon:2005ki} that only for very small values of $W_0$ can
perturbative corrections to the K\"ahler potential be neglected. It was the insight of 
\cite{Balasubramanian:2005zx} that even if $W_0$ is $\CO(1)$ (which is 
more generic than the tiny value for $W_0$ required in KKLT), there can still be a competition 
between the perturbative and nonperturbative corrections to the potential 
in regions of the K\"ahler cone where large hierarchies between the K\"ahler moduli 
are present. We now review this scenario.


\subsection{Large volume scenario (LVS)}
\label{sec:LVS}

As was shown in \cite{Becker:2002nn}, the no-scale structure 
(and factorization of moduli space) is broken by perturbative $\alpha'$ corrections
to the K\"ahler potential, such as  
\be\label{LVS}
K=-\mathrm{ln}(2S_1)-2\,\mathrm{ln}(\V+\tfrac12 \xi S_1^{3/2})+K_{\rm cs}(U,\Ub)\ , 
\ee
where\footnote{Here $\xi$ differs by a 
factor $(2\pi)^{-3}$ from \cite{Becker:2002nn} because we use the string length 
$l_s=2\pi \sqrt{\al}$.} 
$\xi=-\zeta(3)\chi/2(2\pi)^3$ and $S_1 = {\rm Re} \, S$. 
For large volume $\V$, we see that the perturbative correction 
goes as a power in the volume, 
\be \label{expandxi}
-2\ln ( \V + \tfrac12 \xi S_1^{3/2}) = -2\ln \V - \frac{\xi S_1^{3/2}}{\V} 
+ \ldots\ ,
\ee
which by the discussion in the previous subsection will dominate in the scalar potential if
all K\"ahler moduli are even moderately large. 
Using the superpotential 
\be \label{supo}
W = W_0 + W_{\rm np} = W_0 + \sum_i A_i e^{-a_i T_i}\ ,
\ee
the scalar potential has the structure
\be\label{alcuni}
V&=&V_{\mathrm{np1}}+V_{\mathrm{np2}}+V_3\\
&=& e^{K}\left\lbrace G^{\bar \jmath i} \partial_{\bar \jmath}\bar W_{\rm np} \partial_i\W 
+\left[ G^{\bar \jmath i}
K_{\bar \jmath}\,(\bar{W}_0+\bar{W}_{\mathrm{np}}) \partial_i
\W + c.c.\right] \nonumber \right. \\
&&\left. +\left(G^{\bar \jmath i} K_{\bar \jmath} K_i - 3 \right) |W|^2
\right\rbrace \nonumber \; .
\ee
For concrete calculations we will use the model based on the hypersurface of degree $18$ in 
$\mathbb{P}_{[1,1,1,6,9]}^4$ (see \cite{Candelas:1994hw,Denef:2004dm,Curio:2006ea}
for background information on its topology. Some comments about 
generalizations to other models with arbitrary numbers of K\"ahler moduli
are given in appendix \ref{sec:many}). The defining equation is
\be \label{hyper}
z_1^{18}+z_2^{18}+z_3^{18}+z_4^{3}+z_5^2-18\psi z_1z_2z_3z_4z_5-3\phi 
z_1^6z_2^6z_3^6=0 
\ee
and it has the Hodge numbers $h^{1,1}=2$ and 
$h^{2,1}=272$ (only two of the complex structure moduli $\psi$ and $\phi$
have been made explicit in (\ref{hyper}); moreover, not all of the $272$  survive 
orientifolding). We denote the two K\"ahler moduli by 
$T_b=\tau_b+ib_b$ and $T_s=\tau_s+ib_s$, where $\tau_b$ and $\tau_s$ are the 
volumes of 4-cycles, and the subscripts ``$b$'' and ``$s$'' are 
chosen in anticipation of the fact that one
of the K\"ahler moduli $(\tau_b)$ will be stabilized big, and the other one 
$(\tau_s)$ will be stabilized small. An interesting property of this model is that 
it allows expressing the 2-cycle volumes $t_i$ explicitly as functions of the 
4-cycle volumes $\tau_j$, so that the total 
volume of the manifold can be written directly in terms of 4-cycle volumes, yielding
\be \label{tbts}
\V=\frac{1}{9\sqrt{2}}\left( \tau_b^{3/2}-\tau_s^{3/2}\right)\ , \\ \nonumber 
\tau_b=\frac{(t_s+6t_b)^2}{2}\ , \quad \tau_s=\frac{t_s^2}{2}
\ .
\ee
Following \cite{Conlon:2005ki}, we are interested in minima of the potential
with the peculiar property that one K\"ahler modulus $\tau_b \sim \V^{2/3}$ is stabilized
large and the rest  are relatively small (but still large compared to the string scale), 
\be \label{taulnv}
a \tau_s \sim \ln \V \sim \frac32 \ln \tau_b
\ee
in the case at hand. Thus, we expand the potential 
around large volume, treating $e^{-a \tau_s}$ as being of the same
order as $\V^{-1}$. 
In the end one has to check that the resulting potential indeed leads to a minimum 
consistent with the exponential hierarchy $a \tau_s \sim \ln \V$, so that the 
procedure is self-consistent. Applying this strategy,
the scalar potential at leading order in $1/\V$ becomes\footnote{Here we have already stabilized the
axion $b_s$, i.e.\ solved $\partial V / \partial b_s =0$, which
produces the minus sign in the second term; this is also true with many small moduli 
$\tau_i$. See 
appendix \ref{sec:many} for details. Also note that solving $D_UW = 0 = D_SW$ 
causes the values of $U$ and $S$ at the minimum to depend on the 
K\"ahler moduli. However, this dependence arises either from the nonperturbative terms in the 
superpotential or from the $\alpha'$-correction to the K\"ahler potential. Thus it
would only modify the potential at subleading order in the 
$1/\V$ expansion.}
\be  \label{largeVpot}
V_{\CO(1/\V^3)} = \left( \frac{12\sqrt{2} |A|^2  a^2\sqrt{\tau_s} e^{-2 a \tau_s} }{  \V S_1}
- \frac{2 a |AW_0| \tau_s e^{-a \tau_s} }{ \V^2 S_1} + 
\xi\frac{3  |W_0|^2 \sqrt{S_1}}{ 8 \V^3}\right) e^{K_{\rm cs}}\ .
\ee
From here one can see the existence of the large volume minima rather generally.
By the Dine-Seiberg argument \cite{Dine:1985he}, the scalar potential goes 
 to zero asymptotically in every direction. Along the direction (\ref{taulnv}), for large 
volume the leading term in \eqref{largeVpot} is 
\be
V\;\sim \; V_{\rm np2}\; \propto\; -\frac{\ln\V}{\V^3}\ , 
\ee
which is negative, so the potential $V$ approaches zero from below. For moderately
small values of the volume, $V$ is positive (this is guaranteed if the Euler number
$\chi$ is negative, hence $\xi$ positive), so in between
there is a minimum. This minimum is typically nonsupersymmetric, and because we are no
longer balancing a tree-level versus a nonperturbative term, we can
find minima at large volume --- hence the name
large volume scenario (LVS).\footnote{By
``tree-level'' we intend ``tree-level supergravity'', i.e.\ for the 
purposes of this paper
we call both $\al$ and
$g_s$ corrections ``quantum corrections''.}
To be precise, in flux compactifications
we move in
parameter space by the choice of discrete fluxes, but since $\V$ is
exponentially 
sensitive to parameters like $S_1$, large volume minima appear easy to
achieve also by small changes in flux parameters.
If we allow for very small values of $W_0$
(so that KKLT minima exist at all),
 the above minimum can coexist with the KKLT minimum 
\cite{Balasubramanian:2004uy,Conlon:2005ki}.
 Here, we will allow $W_0$ to take generic values of order one.

The astute reader will have noticed that this argument for 
the existence of the LVS minimum is ``one dimensional'', as it only
takes into account the behavior of the potential along the direction
(\ref{taulnv}). One must of course check minimization with respect to all
K\"ahler moduli. In \cite{Balasubramanian:2005zx} a plausibility 
argument to this effect was 
given, and the existence of the minimum was explicitly checked in the case of the 
$\mathbb{P}_{[1,1,1,6,9]}^4$ model by explicitly minimizing the 
potential (\ref{largeVpot})
with respect to the K\"ahler moduli. 
In doing so, it is convenient to trade the two independent variables 
$\{ \tau_b$,$\tau_s \}$ for $\{ \V$,$\tau_s \}$ so that $\partial_{\tau_s}\V=0$, as then
the last term in \eqref{largeVpot} is independent of $\tau_s$  
(this will be different when we include loop corrections).
Extremizing with respect to $\tau_s$, and defining
\be \label{X}
X&\equiv& A e^{-a\tau_s} \; ,
\ee
one obtains a quadratic equation for $X$,
\be \label{quadratic}
0=\frac{\partial V}{\partial \tau_s}&=&\left( -\frac{6\sqrt{2} a^2}{\sqrt{\tau_s}S_1 \V}
\left(4a \tau_s-1 \right)X^2 +\frac{2a|W_0|}{S_1 \V^2}
\left(a \tau_s -1\right)X \right) e^{K_{\rm cs}}\ .
\ee
In (\ref{X}), we chose $A$ to be real as a potential phase can be absorbed 
into a shift of the axion $b$ and disappears after minimization with 
respect to $b$ (see section \ref{sec:many}). Two comments are in order.
The quadratic equation (\ref{quadratic}) has just one meaningful solution 
($X=0$ corresponds to $\tau_s=\infty$). Moreover, when expanding (\ref{quadratic}) 
in $1/(a\tau_s)$, the leading terms arise from derivatives of the exponential. 

Formula (\ref{quadratic}) is an implicit equation determining $\tau_{\rm s}$.
However, one can easily solve \eqref{quadratic} for $X$ and obtains
\be
\label{Xsol_noE}
X=Ae^{-a \tau_s} =\frac{\sqrt{2}| W_0|}{24 a \V } \sqrt{\tau_s} 
\left(1 - \frac{3}{4 a \tau_s}
  \right) \; + \; \mathcal{O}\!\left(\frac{1}{(a \tau_s)^2} \right)\ .
\ee
The hierarchy (\ref{taulnv}) is obvious in this solution, rendering the 
procedure self-consistent. One also notices that reasonably large values of $\tau_{\rm s}$
(e.g.~35) are not difficult to obtain, if $\V$ is 
stabilized large enough; for example, simply set $a \sim 1$, $A \sim 1$, $W_0 \sim 1$.
We fill in the numerical details, following \cite{Balasubramanian:2005zx},
in appendix \ref{sec:11169} (including some further observations).


\section{String loop corrections to LVS} 

As already emphasized, the $\al$ correction proportional to 
$\xi$ is only one among many corrections
in the string effective action. 
We now consider the effect of string loop corrections on this scenario
and what the regime of validity is for including or 
neglecting those corrections. 
Volume stabilization with string loop corrections but without nonperturbative
effects was considered in \cite{Berg:2005yu}. 

To be precise, the
corrections considered in \cite{Berg:2005yu} were those of
\cite{Berg:2005ja}, that were computed for toroidal ${\cal N}=1$ and ${\cal N}=2$
 orientifolds. Here, we would need the analogous corrections for
smooth  Calabi-Yau orientifolds. Needless to say, 
these are not known. Faced with the fact that the string 
coupling $g_{\rm s}$ is stabilized at a finite
(and typically not terribly small) value, we propose that attempting to 
estimate the corrections based on experience with the toroidal case is
better than arbitrarily discarding them. As we will
see, if our estimates are correct, typically the loop corrections can be neglected,
though there may at least be some regions of parameter 
space where they must be taken into account
(see figure \ref{3dfig}).
(In section \ref{otherclasses}, we will briefly consider ``cousins" of LVS
where they cannot be neglected anywhere in parameter space.) 
Improvement on our guesswork would of course be very desirable.

\subsection{From toroidal orientifolds to Calabi-Yau manifolds}
\label{general}

We would like to make 
an educated guess for the possible form of one-loop corrections in a
general Calabi-Yau orientifold. All we can hope to guess is
the scaling of these corrections with the K\"ahler moduli $T$ and 
the dilaton $S$. The dependence on 
other moduli, like the complex structure moduli $U$, cannot be
determined by the following arguments (even in the toroidal
orientifolds this dependence was quite complicated). 

\begin{figure}[th]
\begin{center}
\psfrag{tb}[tc][bc][1][1]{$\tau_{\rm b}$}
\psfrag{ts}[tc][bc][1][1]{$\tau_{\rm s}$}
\psfrag{D3}[tc][bc][1][1]{D3}
\psfrag{a}[tc][bc][1][1]{{\it a}}
\psfrag{b}[tc][bc][1][1]{{\it b}}
\includegraphics[width=0.5\textwidth]{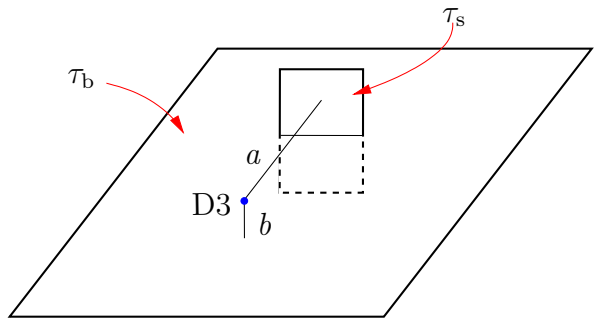}
\end{center}
\vspace{-5mm}
\caption{The loop correction $\Et$ comes from the exchange
of closed strings, or equivalently an open-string one-loop diagram, 
between the D3-brane and D7-branes (or O7-planes)
wrapped on either the small 4-cycle $\tau_{\rm s}$ (as in {\it a}) or the large
4-cycle $\tau_{\rm b}$ (as in {\it b}). The exchanged closed strings carry 
Kaluza-Klein momentum. \label{fig:E3}}
\end{figure}

In order to generalize the results of \cite{Berg:2005ja} to the case of 
smooth Calabi-Yau manifolds, we should first review them and 
in particular remind ourselves where the various
corrections come from in the case of toroidal orientifolds. There,
the K\"ahler potential looks as follows (we will explain the notation
as we go along):
\be \label{Ktorus0}
K&=&-\mathrm{ln}(2S_1)-2\,\mathrm{ln}(\V)+K_{\rm
cs}(U,\Ub) - \frac{\xi S_1^{3/2}} {  \V} \\
&&\hspace{2cm} +\sum_{i=1}^{3}\frac{\Et_i(U,\Ub)}{4\tau_iS_1}
+\sum_{i \neq j \neq k}^3\frac{\Es_k(U,\Ub)}{4\tau_i\tau_j}\ .  \nonumber
\ee   
 There are two kinds of corrections. One comes from the exchange
of Kaluza-Klein (KK) modes between D7-branes (or O7-planes)
and D3-branes (or O3-planes, both localized in the internal space), 
which are usually needed for tadpole cancellation, cf.\ fig.\
\ref{fig:E3}.
This leads to the first kind of corrections in (\ref{Ktorus0}), proportional to
$\Et_i$ where the superscript $(K)$ reminds us
that these terms originate from KK 
modes. In the toroidal orientifold case, this type of correction is 
suppressed by the dilaton and a single K\"ahler modulus $\tau_i$, related to the volume 
of the 4-cycle wrapped by the D7-branes (or O7-planes, respectively).\footnote{We 
should mention that there was no additional correction of this kind coming 
from KK exchange between (parallel) D7-branes in \cite{Berg:2005ja} (actually
that paper considered the T-dual version with D5-branes, but here we 
directly translate the result to the D7-brane language). This was due to the fact that in 
\cite{Berg:2005ja} the D7-brane scalars were set to zero.
In general we would also expect a correction coming from parallel 
(or more generally, non-intersecting) D7-branes by exchange of KK-states.
These should scale in the same way with the K\"ahler
moduli as those arising from the KK exchange between D3- and D7-branes.} 
We expect an analog of these terms to arise more generally, given that they
originate from the exchange of KK states which are present in all 
compactifications. 

\begin{figure}[th]
\begin{center}
\psfrag{tb}[tc][bc][1][1]{$\tau_{\rm b}$}
\psfrag{ts}[tc][bc][1][1]{$\tau_{\rm s}$}
\includegraphics[width=0.5\textwidth]{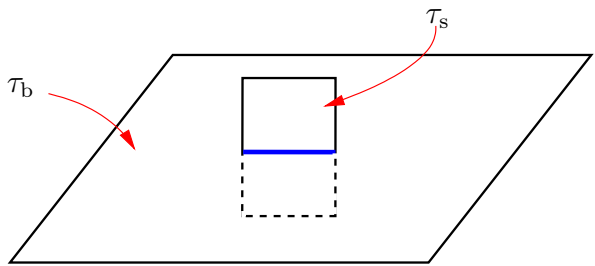}
\end{center}
\vspace{-5mm}
\caption{The loop correction $\Es$ comes from the exchange
of winding strings on the intersection between the
 small 4-cycle $\tau_{\rm s}$ and the large
4-cycle $\tau_b$. If this intersection is empty,
there are no terms with $\Es$. \label{fig:E7}}
\end{figure}

\begin{figure}[th]
\begin{center}
\psfrag{A}[tc][bc][0.8][0.8]{A}
\psfrag{B}[tc][bc][0.8][0.8]{B}
\psfrag{C}[tc][bc][0.8][0.8]{C}
\psfrag{no D-brane}[tc][bc][0.8][0.8]{no D-brane}
\includegraphics[width=0.5\textwidth]{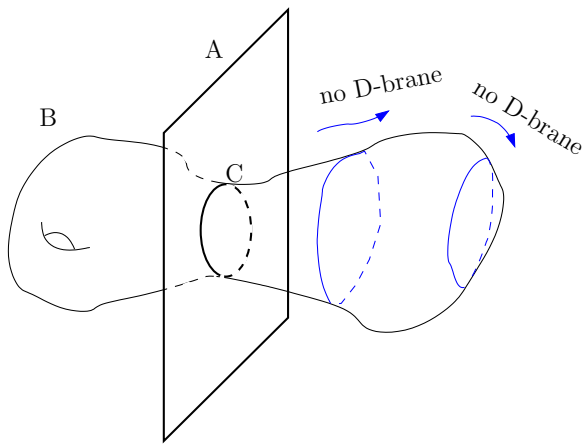}
\end{center}
\vspace{-5mm}
\caption{
A D7-brane is wrapped on a 4-cycle A, which intersects
the 4-cycle B on a 2-cycle C.
For Dirichlet strings, the relevant topological
condition (the existence of
nontrivial 1-cycles) is on the intersection locus C, not on cycle B
or on the whole Calabi-Yau. 
In other words, without the D-brane,
the string on cycle C could have been unwound
by sliding it along cycle B (as shown in the figure).
With the D-brane, the string on cycle C is stuck. \label{fig:wurst}}
\end{figure}

The second type of correction comes from the exchange of winding strings 
between intersecting stacks of D7-branes (or between intersecting 
D7-branes and O7-planes). The exchanged strings are wound around 
non-contractible 1-cycles within the intersection locus of the D7-branes (and O7-planes,
respectively), cf.\ fig.\ \ref{fig:E7}. 
This leads to the second kind of correction in (\ref{Ktorus0}) 
proportional to $\Es_i$. The superscript $(W)$ reminds us that these 
terms arise from the exchange of winding strings. In toroidal orientifolds,
this type of correction is suppressed by the two K\"ahler moduli measuring 
the volumes of the 4-cycles wrapped by the D7-branes (and O7-planes). 
One might a priori think that this kind of correction does not
generalize  easily
to a smooth Calabi-Yau which has vanishing first Betti number (and therefore
at most torsional 1-cycles). However, the exchanged winding strings  
are, from the open string point of view,  
Dirichlet strings with their endpoints stuck on the D7-branes.  Thus,
the topological condition is on the
cycle over which the two D7-brane stacks (or one D7-brane stack and an O7-plane)
intersect, as in figure \ref{fig:wurst}. 
Thus, it depends on the topology
of specific cycles within cycles whether
winding open strings exist in a given model.\footnote{The toroidal 
orientifold case seems to be a bit degenerate. Two stacks of D7-branes 
intersect along a 2-cycle with the topology of $\mathbb{P}^1$. However,
there are point-like curvature singularities along the $\mathbb{P}^1$ at the 
orbifold point and strings winding around these singular points cannot be
contracted without crossing the singularities. This seems to allow 
for stability of winding strings (at least classically).} 

Given the expressions in \cite{Berg:2005ja}
and the subset reproduced in \eqref{Ktorus0} above, it is tempting to conjecture 
that some terms at one loop might be suppressed only by 
powers of single K\"ahler moduli like the $\tau_i$ (and the dilaton):
\be \label{torguess}
\mbox{Calabi-Yau: } \quad \Delta K_{g_{\rm s}} \stackrel{?}{=}  
\frac{\E }{S_1 \tau_i} 
\ee
for some function $\E$ of the complex structure and open string moduli.
If this were the case, the one-loop corrections would typically dominate 
the $\alpha'$ correction in \eqref{Ktorus0}
 (which is suppressed by the overall
volume $\V$) in the K\"ahler potential, if
there are large hierarchies among the K\"ahler moduli. 
However, one should keep in mind that 
toroidal orientifolds are rather special in that they have 
very simple intersection numbers.
In particular, the overall volume can be written as $\cv \sim \tau_i t_i$,
where there is no summation over $i$ implied. Thus, it is not obvious 
whether a generalization
to the case of a general Calabi-Yau really contains terms suppressed 
by single K\"ahler moduli instead 
of the overall volume. Even though we cannot exclude the presence of 
such terms, we deem it
more likely that the scaling of one-loop corrections to the
K\"ahler potential is not \eqref{torguess} but
\be \label{generalize}
\mbox{Calabi-Yau: } \quad \Delta K_{g_{\rm s}} & \stackrel{!}{\sim} &
\frac{\sum_{\rm KK} m^{-2}_{\rm KK}}{S_1  \cv} \sim 
\sum_{\mathfrak{a}} \frac{\gk^{\mathfrak{a}}(t,S_1) \Et_{\mathfrak{a}}}{S_1  \cv}\  \non
& \hspace{-1cm} {\rm and} & 
\frac{\sum_{\rm W} m^{-2}_{\rm W}}{\cv} \sim 
\sum_{\mathfrak{q}} \frac{\gw^{\mathfrak{q}}(t,S_1) \Es_{\mathfrak{q}}}{ \cv}\ , 
\ee
where the sums run over KK and winding states, respectively.
Also, $\Et$ and $\Es$ are again unknown 
functions of the complex structure and open string 
moduli, $t$ stands for the 2-cycle volumes
(in the Einstein frame; see appendix \ref{noscale2a}) and
the functions $\gk(t,S_1)$ and $\gw(t,S_1)$ determine the scaling of the 
KK and winding mode masses with the K\"ahler moduli
and the dilaton.\footnote{In rewriting the sums over 
KK and winding states in terms of the functions $g$ and $\E$, 
we assume that the 
dependence of the corresponding spectra on the complex 
structure and K\"ahler moduli factorizes. In the known examples of
toroidal orientifolds (with or without world-volume fluxes),
this is always the case, cf.\ \cite{Blumenhagen:2006ci}.
Moreover, in general there can appear several contributions
(denoted by $\mathfrak{a}$ and $\mathfrak{q}$) depending 
on which tower of KK or winding states are exchanged 
in a given process. We will see explicit examples of this in the following.}  
As we review in appendix \ref{app:tori}, 
in the toroidal orientifold case 
the suppression by the overall volume arises naturally through the
Weyl rescaling to 
the 4-dimensional Einstein frame.

Starting with the ansatz \eqref{generalize} for smooth Calabi-Yau
manifolds, the known form (\ref{torguess}) 
for toroidal orientifolds follows simply by substituting
 $\gk$, $\gw$ and the intersection numbers for
the toroidal orientifold case.
In particular, $\gk \sim t_i$ for the 2-cycle transverse to the relevant D7-brane, 
while $\gw \sim t_i^{-1}$ for the 2-cycle along which the two D7-branes 
intersect.
Then, the first of the terms in (\ref{generalize}) 
reduces to $\Et_i/(S_1 \tau_i)$ for toroidal 
orientifolds, the second 
to $\Es_i/(\tau_j \tau_k)$ with $j \neq i \neq k$, cf\ \eqref{Ktorus}.
Our strategy in the following chapters 
will therefore be to assume a scaling like (\ref{generalize}) for the 
1-loop corrections to the K\"ahler 
potential for general Calabi-Yau spaces.

As already mentioned,
the dependence on the complex structure and open string moduli cannot
be inferred by analogy to the orientifold case. We parameterize our ignorance
by keeping the expressions $\E$ in
(\ref{generalize}) as unknown functions of the 
corresponding moduli. Then we investigate the consequences of the one-loop terms, 
depending on the 
size of these unknown functions at the minimum of the potential for the 
complex structure and open string moduli. 
Some further comments on the form of $\Delta K_{g_{\rm s}}$ 
will appear in section \ref{toroidal}.


\subsection{LVS with loop corrections}

Thus, allowing for string loop corrections 
of the form \eqref{generalize} in \eqref{LVS},
and expanding the $\alpha'$ correction as in \eqref{expandxi},
we can write
\be\label{LVSLoop}
K&=&-\mathrm{ln}(2S_1)-2\,\mathrm{ln}(\V)+K_{\rm
  cs}(U,\Ub) - \frac{\tilde \xi S_1^{3/2} }{ \V} 
+ \sum_{\mathfrak{a}} \frac{\gk^{\mathfrak{a}} \Et_{\mathfrak{a}}}{ S_1\V}
+\sum_{\mathfrak{q}} \frac {\gw^{\mathfrak{q}} \Es_{\mathfrak{q}}}{\V}\ ,
 \nonumber \\
W &=&  W_0 + \sum_i A_i e^{-a_i T_i}\ ,
\ee
where as explained in the previous section, 
we have not specified the explicit
form of the loop corrections $\E$,
that are allowed to be functions of $U$ (and in general of the open string moduli,
that we neglect in our analysis, assuming that they can be stabilized by fluxes). 
The K\"ahler potential for the complex structure moduli 
$K_{\rm cs}(U,\bar{U})$ is left unspecified in \eqref{LVSLoop}, 
indeed we will not need its explicit form. 
For consistency, we have also included loop corrections to the
$\alpha'$ correction.\footnote{We remind the reader that the $\al$ correction
arises from the $R^4$ term in 10 dimensions whose coefficient  receives
corrections at 1-loop (and from D-instantons). The 1-loop 
correction amounts to a shift of the prefactor 
from $\xi$ to $\tilde{\xi} = \xi \Big(1 + 
\frac{\pi^2}{3 \zeta (3) S_1^2} \Big)$, see for instance 
\cite{Green:1999qt} for a review.} This 
changes $\xi$ to $\tilde{\xi}$, which is a small change; for $S_1=10$,
numerically  $\tilde{\xi} \approx 1.02\, \xi$.

Neglecting fluxes, the functions $\gk^{\mathfrak{a}}$ 
and $\gw^{\mathfrak{q}}$ are 
proportional and inversely proportional to some 2-cycle volume,
respectively. (We will come back to 
corrections from fluxes in appendix \ref{E7neq0}.) 
When using a particular basis of 2-cycles (with volumes $t_i$ as 
in appendix \ref{noscale2a}), the 2-cycle volume appearing 
in $\gk^{\mathfrak{a}}$ or $\gw^{\mathfrak{q}}$ might 
be given by a linear combination $t_{\mathfrak{a}} = \sum_i c_i t_i$ 
of the basis cycles $t_i$ (and
similarly for $t_{\mathfrak{q}}$). Depending on which  2-cycle
is the relevant one,
this linear combination might or might not contain the large 2-cycle 
$t_b\sim \V^{1/3}$, which always exists in LVS. If it is present in the 
linear combination, one can neglect 
the contribution of the small 2-cycles to leading order in a large volume 
expansion and obtains possible terms proportional to $\Et_b S_1^{-1} \V^{-2/3}$ 
or $\Es_b \V^{-4/3}$, where the subscript $b$
refers to the large 4-cycle $\tau_b$. 

Before getting into the details, it is hard to resist 
trying to anticipate what might happen.
For those terms that are {\it more} suppressed in volume than the $\tilde \xi$
term (e.g $\Es_b$), 
one would expect the loop corrections to have little effect on
stabilization. They could still represent a small but interesting
correction to physical quantities in LVS.
For those that are {\it less} suppressed in 
volume than the $\tilde \xi$
term (e.g. $\Et_b$), one would expect the loop correction to have a huge effect on
stabilization, and severely constrain the allowed values for the 
complex structure moduli and the dilaton in LVS (in particular, 
constrain them to a region in moduli space where the function
$\Et_b$ takes very small values). 
We will find, however, that this expectation is sometimes too naive.
For example, there can be cancellations in the scalar potential
that are not obvious from just looking at the K\"ahler potential. 

Let us now get into more detail on
what happens in the LVS model with loop corrections.


\subsection{The $\mathbb{P}_{[1,1,1,6,9]}^4$ model} \label{P model}

We would now like to specify the general form of the K\"ahler-
and superpotential \eqref{LVSLoop} to the case of the 
$\mathbb{P}_{[1,1,1,6,9]}^4$ model.
In this space,
the divisors that produce
 nonperturbative superpotentials when D7- (or D3-) branes are wrapped around
them do not intersect, as reviewed for instance in 
\cite{Curio:2006ea}. 
Therefore, we do not expect any correction of the 
$\Es$ type in this model (for the generalization
to models where there are such intersections, see appendix \ref{E7neq0}).  
Moreover, we neglect flux corrections to the KK mass spectrum 
in the main text. It is shown in appendix \ref{E7neq0} that, for small 
fluxes, this correctly
captures all the qualitative features we are interested in, 
and it leads to much clearer formulas. 
Thus, we now consider the scalar potential resulting from
\be \label{realistic k}
K&=&-\mathrm{ln}(2S_1)-2\,\mathrm{ln}(\V)+K_{\rm
  cs}(U,\bar U) - \frac{\tilde{\xi} S_1^{3/2}}{  \V}
  +\frac{\sqrt{\tau_b} \Et_b }{ S_1\V} 
  + \frac{\sqrt{\tau_s} \Et_s }{ S_1\V}\ ,
  \nonumber \\ [1mm]
W &=&  W_0 +  A e^{-a T_s}.
\ee 
As $\tau_b$ is very large the corresponding non-perturbative term in
the superpotential of \eqref{LVSLoop} can be neglected, which allowed us 
to simplify the notation by setting 
$A_s=A$ and $a_s = a$. 

The general structure of the scalar potential was already given in 
(\ref{alcuni}).
The three contributions at leading order (${\cal O} (\V^{-3})$) 
in the large volume expansion are
\be \label{full loop}
V_{\mathrm{np1}}&=&e^{K_{\rm cs}} \frac{24 a^2|A|^2\tau_s^{3/2}
e^{-2a\tau_s}} {\V\Delta}\ ,\\[0.5cm]
V_{\mathrm{np2}}&=&-e^{K_{\rm cs}}  \frac{2a|AW_0|\tau_s
  e^{-a\tau_s}}{ S_1 \V^2}\left[ 1+ \frac{6\Et_s}{\Delta} \right]\ ,\\[0.5cm]
V_3&=& \frac{3 e^{K_{\rm
      cs}}|W_0|^2 } { 8 \V^3}
      \left[
S_1^{1/2} \tilde{\xi} + \frac{4(\Et_s)^2\sqrt{\tau_s} }{
S_1^{2}\Delta}\right]\ , \label{Vwnpzero}
\ee
where the axion has already been minimized for, as discussed in section \ref{sec:many}, and 
\be \label{Delta}
\Delta&\equiv& \sqrt{2} S_1\tau_s-3\Et_s\ .
\ee
The leading $\al$-correction 
is the $\tilde{\xi}$ term in $V_3$ above.
We now see that it scales with the volume and the 
string coupling $g_{\rm s}=1/S_1$ 
as claimed in the Introduction, in
eq.\ (\ref{orders3}). Also the volume dependence of the 
loop correction ($\Et_s$ term) in $V_3$
 is as announced in (\ref{orders3}). The 
$g_{\rm s}$ factors seem to differ from \eqref{orders3}; we see
$g_s^2$, $g_s^2$ and $g_s^3$ for 
$V_{\mathrm{np1}}$, $V_{\mathrm{np2}}$ and $V_3$, respectively.
This is because the 
$g_{\rm s}$ dependence advertised 
in (\ref{orders3}) arises in models where, unlike
in  $\mathbb{P}_{[1,1,1,6,9]}^4$,
 the $\Es$
correction is present as well, cf.\ appendix \ref{E7neq0}.\footnote{There, it 
is shown that including the effect of fluxes on the KK spectrum 
might also produce this behavior.}
It is also worth mentioning that
the loop correction
proportional to $\Et_s$ modifies $V_{\mathrm{np1}}$ and 
$V_{\mathrm{np2}}$ at leading order in the $\V$-expansion.
whereas the $\al$ correction
does not; it only appears in $V_3$. This is so even though both corrections are 
equally suppressed in the K\"ahler potential (i.e.\ $\sim \V^{-1}$). The
reason for this can be  traced back to the fact that the
loop-correction
explicitly depends 
 on $\tau_s$ and not only on the overall volume, cf.\ the
discussion in appendix \ref{canc} and \ref{cancnp}. 

\begin{figure}[th] \begin{center}
\includegraphics[width=0.4\textwidth]{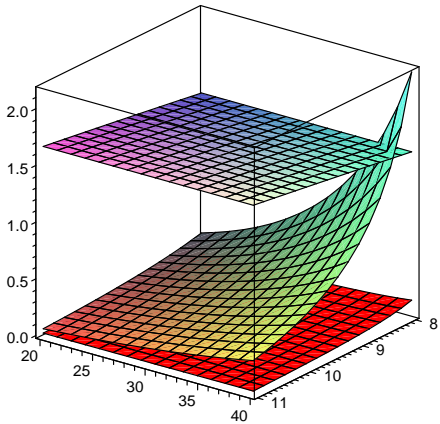}
\end{center}
\vspace{-5mm}
\caption{The top surface is the $\alpha'$ correction, the second is the $g_{\rm
  s}$ correction, and the ``red carpet'' is $10/\Delta$ (we used the values
  $A=1, W_0=1, a=2 \pi/8$).
    \label{3dfig}
We see that for most of the parameter range,
the $\alpha'$ correction dominates, and only for
large $\Et_s$, with the string coupling $g_{\rm s}=1/S_1$ not
too small, do
the contributions become comparable. \label{fig:compare}}
\end{figure}

As anticipated, $\Et_b$ and its first derivatives appear only at the
next order, $\CO(\V^{-10/3})$:
\be\label{e31}
V_{10/3}=2\frac{6^{1/3} |W_0|^2 e^{K_{\rm cs} } } {S_1^3\V^{10/3}} \left[ (\Et_b)^2 
+\frac{3}{4} \partial_{\alpha} \Et_b \partial_{\bar \alpha}\Et_b 
K_{\rm cs}^{\alpha \bar \alpha} \right],
\ee
where $\partial_{\alpha}=\partial/\partial U^{\alpha}$ and
$\partial_{\bar \alpha}=\partial/\partial \bar{U}^{\bar \alpha}$ 
and $\alpha$ enumerates the complex structure moduli. 
For $\Et_b=\Et_s=0$, the potential terms at leading order 
coincide with the original case discussed in \eqref{potte}, cf.\ appendix 
\ref{sec:11169}.
The singularity from zeros of the denominator is an artifact of the expansion as 
discussed in appendix \ref{loopcorrected}. The range of 
validity is limited to the range in moduli space
where the denominator $\Delta$ does not become too small. It is also apparent
that the loop terms are subleading in a large $\tau_s$, large $S_1$
expansion. However, depending on the relative values of the parameters
$\{ \Et_s,\tau_s,S_1\}$, a truncation to the first terms in such an
expansion may or may not be valid. We perform a numerical comparison
of the two contributions to $V_3$ in figure \ref{fig:compare}.  

We can understand the volume dependence of the terms 
\eqref{full loop}-\eqref{e31} as follows. The common prefactor 
$e^{K}$ gives an overall suppression $\tau_b^{-3}\simeq\V^{-2}$. 
The quantum corrections obey 
the rule that a term proportional to $1/\tau_b^{\lambda}$ in $K$ appears in $V_3$ 
at order $1/\tau_b^{\lambda+3}$ (where the $+3$ comes from the 
overall $e^{K}$ factor) for all values of $\lambda$ \textit{except} 
for $\lambda = 1$. 
When it does appear, it is generated 
by the term $(K^iK_i-3)$ and breaks the no-scale structure. For 
$\lambda=1$ there is a cancellation at leading order, 
so it appears only at order $1/\tau_b^{2+3}$ 
(see appendix \ref{canc ap} and \ref{canc}).
This rule can explicitly be verified in our calculation: the $\al$ 
and  the $\Et_s$ corrections are suppressed by $1/\tau_b^{3/2}$ in $K$, and 
therefore they appear with the suppression $1/\tau_b^{9/2}$ in $V_3$. 
On the other hand, for the $\Et_b$ term a cancellation takes place 
to leading order ($\lambda =1$). It appears neither 
in $V_{\mathrm{np1}}$ nor in $V_{\mathrm{np2}}$ at leading order (which 
can be understood more generally, cf.\ appendix \ref{cancnp}). Thus, 
it only appears subleading
in the potential, at $\mathcal{O}(\V^{-10/3})$.\footnote{This 
cancellation for $\lambda = 1$ was already noticed 
in \cite{Berg:2005yu}, albeit in the case without nonperturbative superpotential. 
In \cite{Choi:2007ka} it was argued that this cancellation can be understood from 
a field redefinition argument combined with the no-scale 
structure of the tree-level K\"ahler potential. 
That argument holds for the case of a single K\"ahler modulus $T$ 
with tree-level K\"ahler potential $-3 \ln(T+\bar T)$ and under the assumption 
that the coefficient of the loop correction to the K\"ahler potential 
$\sim (T+\bar T)^{-1}$ is independent of the complex 
structure moduli and the dilaton. Here, these assumptions do not hold, 
but we showed that the term $\sim (T_b+\bar T_b)^{-1}$ in 
the K\"ahler potential nevertheless 
only appears at subleading order in the \textit{potential}
in LVS, cf.\ (\ref{full loop})-(\ref{e31}). \label{cancelfn}}

We now proceed to  minimize the potential (\ref{full loop})-(\ref{Vwnpzero}), 
using the same 
strategy as in
the case without loop corrections, cf.\ section \ref{sec:LVS} 
and appendix \ref{sec:11169}. 
The equations $\partial_\V V=0=\partial_{\tau_s}V$ are of course more complicated now,
 but it is easy to solve them numerically. Doing so we find 
that the volume $\V$ and the small 4-cycle volume
$\tau_s$, viewed as functions of $S_1$ and $\Et_s$, 
are well fit by linear functions when restricted to a sufficiently limited range 
in parameter space. For example,
\be \label{fits}
  \hspace{-4cm}\underline{ \mbox{range: } S_1 = [8,11],\,  \Et_s=[20,40]}\hspace{-2cm}
&& \nonumber \\[1mm]
 \log_{10} \V &=& 1.720 \, S_1 -0.1208\, \Et_s -3.437\ ,   \\
 \tau_s &=& 5.000 \, S_1  -0.3581\,  \Et_s -8.638\ .    \nonumber 
\ee
The fits are quite good; the error is no greater than $\pm 0.3$
for $\tau_s$ and $\pm 0.1$ for $\log_{10}\V$ in this range,
for an $\{S_1,\Et_s\}$ grid of $40^2$ points. 

From (\ref{fits}) we see an interesting difference to the case 
without loop corrections. The 
value of $\tau_s$ at the minimum depends on the complex structure
moduli $U$, through $\Et_s$. This is in contrast to the case 
without loop corrections, where the value of $\tau_s$ is only determined 
by the value of the Euler number $\xi$ and the dilaton $S_1$, cf.\
(\ref{sol}) below. It is analogous to the perturbative stabilization
in \cite{Berg:2005yu} where the volume at the minimum of the potential
also depends on $U$.

The result (\ref{full loop})-(\ref{Vwnpzero}) was derived in a particular 
model, but we expect the appearance of loop corrections in $V$ 
to be more general. This opens up the possibility 
that in principle, one might obtain large volume minima even 
for manifolds of vanishing (or even positive) Euler number, 
where LVS is not applicable, as LVS-style stabilization 
only holds for one sign of $\xi$. In practice it might be difficult 
to get large enough values for the 1-loop 
corrections to stabilize $\tau_s$ at a value sufficiently bigger than 
the string scale. This deserves further study. 

We also note that the special structure of (\ref{Vwnpzero}) and (\ref{e31}), 
i.e.\ the appearance of $\E_s$ only in (\ref{Vwnpzero}) and of $\E_b$ only 
in (\ref{e31}), offers additional flexibility in tuning the relative size 
of these terms in a purely perturbative stabilization of the 
K\"ahler moduli along the lines of \cite{vonGersdorff:2005bf,Berg:2005yu}. 
Also this point deserves further study.


\section{Gaugino masses}
\label{sec:gaugino}

Now that we know how the stabilization of the (K\"ahler) moduli 
is modified by loop corrections, 
it is natural to extend our analysis to the soft supersymmetry
breaking Lagrangian (For 
a review see for instance \cite{Nilles:1983ge,Martin:1997ns}.)
In LVS, supersymmetry breaking
is mostly due to $F$-terms: 
$F_s\neq0$, $F_b\neq 0$. These determine the soft supersymmetry breaking 
terms which can be present in the low energy effective action 
without spoiling the hierarchy between the electroweak and the Planck scale,
\be
\mathcal{L}_{\rm eff}=\mathcal{L}_{\rm MSSM}+\mathcal{L}_{\rm soft}\ .
\ee
The soft Lagrangian contains gaugino masses $M$, scalar masses $m$,
further scalar bilinear terms $B$ and trilinear terms $A$. (For 
explicit expressions, see the aforementioned reviews, or e.g.\ \cite{Conlon:2006us}.)

Let us start considering gaugino masses. In \cite{Conlon:2006us} 
it was shown that in LVS, 
gaugino masses $M_a$
are generically suppressed with respect to the gravitino mass $m_{3/2}$:
\be 
|M_a| \simeq \frac{m_{3/2}} {\mathrm{ln} (1/m_{3/2})}  
\left[ 1+\mathcal{O}\left( \frac{1} {\mathrm{ln}
(1/m_{3/2} ) }\right)\right]
\ee
(we use units in which $M_{\rm Pl}=1$). 
This suppression results from a cancellation of the leading order
$F$-term contribution to gaugino masses. 
We briefly review this calculation. 
Given the $F$-terms
\be \label{Fterms}
F^I = e^{K/2} G^{\bar J I} D_{\bar J} \bar W \; ,
\ee
gaugino masses are given by \cite{Nilles:1983ge}
\be 
\label{Ma_general}
M_a=\frac{1}{2}\frac{1}{{\rm Re} f_a} \sum_{I}F^{I}\partial_{I}f_a\ ,
\ee
where $f_a$ are the gauge kinetic functions and $a$ labels
the different gauge group factors. In LVS the Standard Model (SM) gauge groups 
arise from D7-branes wrapped around small 4-cycles. We do not try 
to go into the details of how to embed the SM concretely, but we 
mention that different gauge group factors might arise from 
brane stacks wrapping the same 4-cycle if world volume fluxes are 
present on the branes. In that case the gauge kinetic functions 
are given by\footnote{We use the ``phenomenology'' normalization
of the gauge generators, in the language of \cite{Kaplunovsky:1987rp}; 
that explains the relative factor of $4 \pi$ in \eqref{gaugekinetic}.}
\be \label{gaugekinetic}
f_a= \frac{T_a}{4 \pi} + h_a({\cal F}) S + f_a^{(1)}(U)\ ,
\ee
where $h_a$ depends on the world volume fluxes and we also included
a possible 1-loop correction to the gauge kinetic function which depends 
on the complex structure (and possibly open string) moduli. If 
several gauge groups arise from branes wrapped around the same cycle, 
the same K\"ahler modulus $T$ would appear in all of them. 
From (\ref{gaugekinetic}) it is clear why the D7-branes of the SM have 
to wrap small 4-cycles, because otherwise the gauge coupling would
come out 
too small (unless there is an unnatural cancellation between the 
different contributions to $f_a$).

As is also apparent from (\ref{gaugekinetic}), the gauge kinetic function 
in general depends not only on the K\"ahler moduli but also on the dilaton and 
the complex structure. Thus, according to (\ref{Ma_general}) we need to know 
$F^U, F^S$ and $F^i$ for the small K\"ahler moduli.\footnote{With a slight abuse 
of notation, we denote the $F$-terms of the K\"ahler moduli by the index $i$, 
but the $F$-terms of the other moduli are identified by the symbol for 
the corresponding modulus, like $F^S$.
This is to avoid introducing too many  indices.}
From the definition \eqref{Fterms},
it is clear that $F^U$ and $F^S$ might be non-vanishing even though 
we assume $D_UW = 0 = D_S W$, provided the 
inverse metric $G^{\bar J I}$ contains mixed components between 
K\"ahler moduli on the one hand and complex structure moduli 
and dilaton on the other hand. Without loop corrections (i.e.\
considering only the leading $\al$ correction) there is no mixing 
between the K\"ahler and complex structure moduli, and one 
finds
\be \label{Fwithout}
F^U = 0 \; , \quad F^S \sim {\cal O}(\V^{-2}) \quad {\rm and} \quad
F^i \sim {\cal O}(\V^{-1}) \quad {\rm (without \ loop\  corrections)}\ .
\ee
Thus, at leading order in the large volume expansion, the sum in 
(\ref{Ma_general}) only runs over the K\"ahler moduli. Moreover, 
taking into account the linear dependence of the gauge 
kinetic functions (\ref{gaugekinetic}) on the (small) K\"ahler moduli, 
the sum effectively only involves a single term, i.e.\
\be
M_a=\frac{1}{8 \pi} \frac{1}{{\rm Re} f_a} F^{a} + {\cal O}(\V^{-2})\ ,
\ee
where $F^a$ is the F-term of the (small) K\"ahler modulus appearing in $f_a$. 

As a concrete example we consider again the $\mathbb{P}_{[1,1,1,6,9]}^4$ model
with only one small K\"ahler modulus $\tau_s$. The corresponding 
F-term is given by  
\be \label{fs}
F^s & = & e^{K/2} \left( G^{\bar s s} \partial_{\bar s} \bar W 
+ (G^{\bar s s} K_{\bar s} + G^{\bar b s} K_{\bar b}) \bar W \right) \nonumber \\
& = & 2 \tau_s e^{K/2} \bar W_0 \left( \Big(1 - \frac{3}{4 a \tau_s}\Big) - 1 
+{\cal O}((a \tau_s)^{-2}) \right) + {\cal O}(\V^{-2}) \ ,
\ee
where we used (\ref{Xsol_noE}) and \eqref{axionmin}
for the first term and (\ref{GK}) for the second. 

Now the leading order cancellation is obvious in (\ref{fs}). 
Determining
the gaugino masses requires dividing by ${\rm Re} f_s$, cf.\ 
(\ref{Ma_general}). In order to further evaluate this,
\cite{Conlon:2006us,Conlon:2006wz} assumed that the 
dilute flux approximation $f_s=(4 \pi)^{-1} T_s$ is valid, i.e.\ they 
neglected the contributions from world-volume fluxes and one-loop
terms compared to the tree-level term. This puts some constraints on the 
allowed discrete flux values determining $h_s$. We want to stress 
that the cancellation appearing in \eqref{fs} is independent 
of this approximation. We are mainly interested in the fate of 
this cancellation when including loop corrections, and do not have 
anything to add concerning phenomenological constraints 
that may arise from imposing the dilute flux approximation. 
Using it, the gaugino masses simplify to
\be \label{gaugino expl}
|M_s|&=&\left| \frac{F^s}{2\tau_s} \right|
= e^{K/2} |W_0| \left(\frac{3}{4 a \tau_s}
+ {\cal O}((a \tau_s)^{-2}) \right) \\[1mm]
&\sim& \frac{m_{3/2}}{\mathrm{ln}(1/m_{3/2})}
\left[1+\mathcal{O}\left(\frac{1} 
{{\rm ln}(1/m_{3/2}) } \right) \right]\ ,  \nonumber
\ee
which is the announced result. In \eqref{gaugino expl} we used
\be \label{m32}
m_{3/2} \sim |W_0| / \V  \qquad {\rm and} \qquad 
a \tau_s \sim \ln (\V / |W_0|)\ ,
\ee
where the second relation holds in LVS due to \eqref{Xsol_noE}.


\subsection{Including loop corrections}

The previous section was  a review of the results found in \cite{Conlon:2006us}. 
Now we ask what changes if one considers 
the loop corrected K\"ahler potential \eqref{realistic k}. 
A priori, as \eqref{gaugino expl} results from a leading order cancellation,
one might wonder whether loop corrections might spoil this small hierarchy 
between the gaugino and gravitino masses. 
To address this concern we start by observing that the gaugino masses 
are still determined by the F-terms of the small K\"ahler moduli (in the
large volume limit). More precisely, the scaling of the F-terms \eqref{Fwithout}
now becomes 
\be \label{Fwith}
F^U = \; {\cal O}(\V^{-2}), \quad F^S \sim {\cal O}(\V^{-2}) \quad {\rm and} \quad
F^i \sim {\cal O}(\V^{-1}) \quad {\rm (with \ loop\  corrections)}\ ,
\ee
i.e.\ $F^U$ no longer vanishes, but it is just as suppressed as $F^S$.

We again focus on the $\mathbb{P}_{[1,1,1,6,9]}^4$ model and ask how 
\eqref{fs} is modified by loop corrections. Amongst other things,
we need to generalize equation \eqref{Xsol_noE} to include loop corrections, 
since we need it to calculate the first term in \eqref{fs}. 
Thus, we need to extremize the potential again with respect to $\tau_s$ by setting
%
%
%
\be \label{dtauV}
\partial_{\tau_s} V &=& \Bigg\{ -\frac{12 \sqrt{\tau_s} a^2}{\V \Delta^2} 
\Big[\Big(4 a \tau_s -1\Big) \Delta + 6 \Et_s \Big] X^2 \nonumber \\
&&  \mbox{} + \frac{2 a |W_0|}{\V^2 S_1 \Delta^2} \Big[ 
\Big(a \tau_s -1\Big) \Big(\Delta^2 - 18 (\Et_s)^2\Big)
+ 6 \sqrt{2} a S_1 \tau_s^2 \Et_s \Big] X \\
&& -\, \frac{3 |W_0|^2 (\Et_s)^2}{4 S_1^2 \V^3 \Delta \sqrt{\tau_s}} \,
\Bigg\}\; e^{K_{\rm cs}} \nonumber
\ee
to zero. Obviously, $X=0$ is no longer a solution. Instead, there are 
 now two non-trivial solutions, one of which goes to zero in the limit 
$\Et_s \rightarrow 0$. This solution corresponds to a maximum of the potential,
so it is of no use to us here. 
We can expand the other solution for large $a\tau_{\rm s}$,
as in the case without loop corrections, yielding
\be
\label{Xsol_withE}
X=Ae^{-a \tau_s} =\frac{\sqrt{2}| W_0|}{24 a \V } \sqrt{\tau_s} 
\left(1 - \frac{3}{4 a \tau_s}\left(1-\frac{2\sqrt{2}a\Et_s  } { S_1}\right)
  \right) + \mathcal{O}\!\left(\frac{1}{(a \tau_s)^2} \right)\ .
\ee
Another ingredient we need is the quantity $G^{\bar \imath s}K_{\bar \imath}$,
in order to evaluate the second term in \eqref{fs}. 
Using equation \eqref{inv metric} we obtain
\be \label{GK_withE}
G^{\bar \imath s}K_{\bar \imath}&=&-2\tau_s \Big(1 + \frac{6 \Et_s}{\Delta} 
\Big)+ \ldots \nonumber \\
&=&-2\tau_s-\frac{6 \sqrt{2} \Et_s}{S_1}-\frac{18(\Et_s)^2}{S_1^2\tau_s}
-\frac{27\sqrt{2} (\Et_s)^3}{\tau_s^2 S_1^3}
+\mathcal{O}\left( \frac{1}{\tau_s^3}\right)+\ldots\ ,
\ee 
where the ellipsis represents terms that are more suppressed in $\V^{-1}$.

Now we see from \eqref{Xsol_withE}, \eqref{GK_withE} and \eqref{inv metric}
that 
 at leading order in an expansion in  $a \tau_s$,
the quantities relevant to evaluate \eqref{fs} are not affected 
by the loop corrections. Thus,
the leading order cancellation in the gaugino masses survives the inclusion 
of loop effects.\footnote{One might argue that this result was to be expected,
because the main assumption of \cite{Conlon:2006us} 
is that the stabilization is due 
to nonperturbative effects, i.e.\ the dominant effect 
in $\partial_{\tau_s} V $ should arise
from the nonperturbative superpotential. However, 
in view of \eqref{dtauV}, it is
no longer obvious that the nonperturbative 
terms dominate when loop corrections are included.} 
At first glance, though, equations
\eqref{Xsol_withE}, \eqref{GK_withE} 
and \eqref{inv metric} seem to 
suggest a correction to the subleading term, that could potentially 
give a significant contribution to the gaugino masses  after the
leading-order
cancellation, cf.\ \eqref{fs}.

In the actual calculation,
this contribution drops out.
 Putting all the ingredients together (and employing the 
dilute flux approximation again), the gaugino mass 
turns out to be
\be  \label{subsub}
|M_s|&=&\left| \frac{F^s}{2\tau_s} \right|
=3 e^{K/2}|W_0| \left|\frac{1}{4a\tau_s}
+\frac{1}{16a^2\tau_s^2}+\frac{S_1-12\sqrt{2}a\Et_s}{64S_1a^3\tau_s^3}
+\ldots \right| \nonumber \\
&\sim &\frac{m_{3/2}}{{\rm ln}(1/m_{3/2})}\left[ 1+\mathcal{O}\left(
\frac{1}{{\rm ln}(1/m_{3/2}) } \right)\right]\ .
\ee
The result of \cite{Conlon:2006us} is therefore very robust. Unexpectedly, 
the correction 
to \eqref{gaugino expl} due to $\Et_s$ only appears at sub-sub-leading order 
in the $1/\ln(1/m_{3/2})$ expansion. 


\subsection{Other soft terms}
\label{sec:othersoft}

In \cite{Conlon:2006wz} all other soft terms were calculated for LVS. 
The main result (see p.\ 15 of \cite{Conlon:2006wz}) is that roughly speaking,
all the soft parameters are determined by $F^s$ and by the 
power with which the chiral matter metrics scale with $\tau_s$. As we 
saw in the previous section, $F^s$ gets modified by loop corrections only 
at sub-sub-leading order in a $1/\tau_s$ expansion (see \eqref{subsub}). Therefore, 
the calculation of all the soft terms in \cite{Conlon:2006wz} 
appears to be quite robust against including loop effects.

One of the key assumptions in 
\cite{Conlon:2006wz} is that all the Yukawa couplings $Y$ 
are already present in perturbation theory, i.e.\ they have the 
schematic form $Y=Y^{\rm pert}(U)+Y^{\rm np}(e^{-T})$. This requirement 
featured prominently  already  in the derivation of the volume dependence
of the chiral matter metrics in \cite{Conlon:2006tj} by scaling arguments. 
In \cite{Conlon:2006wz} 
the same schematic form is essential for simplifying the trilinear soft 
terms $A$. In general these terms receive a contribution of the schematic form
\be \label{Atrems}
F^{T}\partial_T {\rm log} Y=
F^T\frac{\partial_T(Y^{\rm pert}(U)+ 
Y^{\rm np}(e^{-T})) } 
{Y^{\rm pert}(U)
+Y^{\rm np}(e^{-T})}
\sim \frac{\CO (e^{-T})} 
{\CO (T^0)
+\CO (e^{-T})}\ ,
\ee
which is exponentially suppressed if and only if 
$Y^{\rm pert}(U)$ is non-vanishing. However, in 
many examples the Yukawa couplings are actually only generated 
nonperturbatively, see for instance the discussion in 
\cite{Berenstein:2006aj}, and \cite{Ibanez:2001nd} for some examples.
This poses a constraint on the way the Standard Model is 
realized in LVS, if one wants to ensure flavor universality of the 
soft breaking terms as advertised in \cite{Conlon:2006wz}.

One more comment about the important issue
of flavor universality. 
In \cite{Conlon:2006wz}, section 3.4.,
it was argued that in LVS, approximate flavor universality
is a natural consequence of the zeroth-order
factorization of K\"ahler and complex structure moduli 
spaces. 
We provide some more details on the factorized approximation
in appendix \ref{calcu}.


\section{LVS for other classes of Calabi-Yau manifolds?}
\label{otherclasses}

In section \ref{P model} and 
\ref{sec:gaugino} we saw that the 1-loop corrections to the 
moduli K\"ahler potential only have relatively small effects on 
the large volume scenario based on the $\mathbb{P}_{[1,1,1,6,9]}^4$ 
model of \cite{Balasubramanian:2005zx}. In this chapter, we would like 
to ask the question how generic the ``Swiss cheese'' form is for a 
Calabi-Yau manifold and if there are other models in which
the one-loop corrections discussed above might become more important.
This is indeed to be expected if the Calabi-Yau under consideration 
has a fibered structure, as we explain in the following.
If $g_{\rm s}$ corrections do dominate $\alpha'$ corrections, 
they could ruin the volume expansion of LVS.


\subsection{Abundance of ``Swiss cheese'' Calabi-Yau manifolds}
\label{sec:swiss}

In the LVS examples discussed in 
\cite{Conlon:2005ki} the volume in terms of the K\"ahler moduli takes the 
``Swiss cheese'' form 
\be \label{volume}
\V = \left(\tau_b + \sum a_i \tau_{i} \right)^{3/2} - \left( \sum b_i \tau_{i}
\right)^{3/2} -\ldots - \left( \sum c_i \tau_{i} \right)^{3/2}\ ,
\ee
where the coefficients $a_i, ... , c_i$ are only non-vanishing 
for the small K\"ahler moduli. 
The LVS limit consists in 
scaling the overall volume of the Calabi-Yau more or less 
isotropically while having small holes inside the 
manifold. The $\tau$'s are linear combinations
of $\partial_{t_i} \V$, 
where now $\V$ is considered as a (cubic) 
function of the 2-cycle volumes $t_i$. 
For the effective field theory analysis
to be valid one should not only demand that the 4-cycle volumes $\tau_i$ 
are large compared to the string scale, but also that
the 2-cycle volumes 
$t_i$ are large.
In the cases discussed in
\cite{Conlon:2005ki}, the linear combinations $\partial_{t_i}\V$
are indeed such that one can 
have one of them exponentially 
large and the others small (but still sufficiently larger than the string scale), 
without taking any of the $t_i$ to be 
exponentially small. This is obvious for the $\mathbb{P}_{[1,1,1,6,9]}^4$
example where the 2-cycle volume $t_b$ only appears in the definition 
of one of the $\tau$'s, cf.\ (\ref{tbts}), but it is also true for the 
second example of \cite{Conlon:2005ki}, cf.\ their formulas (84). 

However, the $\cf_{18}$ model of \cite{Denef:2004dm} does not seem to 
allow its volume to be written in the form (\ref{volume}) 
with one K\"ahler modulus $\tau_b$ 
that can become large while keeping all the others small (again,
demanding that the $t_i$ stay larger than $1$ in string units).
Thus, it is an interesting question how generic or non-generic the 
``Swiss cheese'' Calabi-Yau manifolds are. We do not attempt to give a
general answer; instead, 
we turn to two examples in which the form of the volume 
differs from (\ref{volume}). 


\subsection{Toroidal orientifolds}
\label{toroidal}

The reason loop corrections
may be more important in toroidal orientifolds than
in compactifications on ``Swiss cheese'' 
Calabi-Yau manifolds is the following. As we already 
discussed in section \ref{general}, the conjectured form of 
1-loop corrections (\ref{generalize}) simplifies in the case 
of toroidal orientifolds, because they have very special and simple 
intersection numbers. More concretely, 
using the definition $\tau_i = \partial_{t_i} \V$, together with 
the special form of the intersection 
numbers in the toroidal case, i.e.\ $\V = t_1 t_2 t_3$, 
the volume can alternatively be expressed as 
\be
\V = \sqrt{\tau_1 \tau_2 \tau_3}= t_i \tau_i\quad 
\mbox{(no summation; $i = 1,2$ or 3)}\ .
\ee
Thus, formula (\ref{generalize}) simplifies and the 1-loop corrections 
proportional to $\Et_i$ 
are only suppressed by single K\"ahler moduli instead of by the overall 
volume. Also the terms proportional to $\Es_i$ can be 
rewritten in the toroidal orientifold case and 
the K\"ahler potential takes the form (for the $T^6/(\mathbb{Z}_2 \times 
\mathbb{Z}_2)$ example)
\be \label{Ktorus}
K&=&-\mathrm{ln}(2S_1)-2\,\mathrm{ln}(\V)+K_{\rm
cs}(U,\Ub) - \frac{\xi S_1^{3/2}} {  \V} \\[1mm]
&& \hspace{2cm} + \sum_{i=1}^{3}\frac{\Et_i(U,\Ub)}{4\tau_iS_1}
+\sum_{i \neq j \neq k}^3\frac{\Es_k(U,\Ub)}{4\tau_i\tau_j}\ ,  \nonumber 
\ee
where the functions $\E$ are non-holomorphic Eisenstein series in this case
\cite{Berg:2005ja}. 
It is easy to see that the origin of this simplification is the fact that
there is just a 
single non-vanishing intersection number in the toroidal orientifold case 
and all K\"ahler moduli appear 
linearly in the cubic expression for the volume. 

The difference of the toroidal orientifold to the 
``Swiss cheese'' case of LVS can also be seen in the different forms of 
the functional dependence of the 
volume on the K\"ahler moduli. 
In the toroidal orientifold case
one has the relations 
\be
\partial_{t_1} \V = t_2 t_3 \quad , \quad 
\partial_{t_2} \V = t_1 t_3 \quad , \quad   
\partial_{t_3} \V = t_1 t_2\ ,
\ee
so that  two of them  will always become large if one takes 
one of the $t_i$ to be large and demands that the other two stay 
larger than $1$. This also holds for any linear 
combinations of them. The difference is also obvious from the 
fact that the 2-cycle volume $t_b$ that is responsible for $\tau_b$
becoming large in the LVS examples of \cite{Conlon:2005ki}
always appears cubically in the volume. This is related to the fact that 
the term $(\tau_b + \sum a_i \tau_{i})$ should be the square of a linear 
combination of the $t_i$, in order for (\ref{volume}) to 
be expressible as a cubic polynomial in the $t_i$. 
In contrast, any (untwisted) 
2-cycle volume in the toroidal orientifold case
only appears {\it linearly} in the 
cubic volume polynomial.

To illustrate the effect of terms in the K\"ahler potential 
that are suppressed only by single K\"ahler moduli 
instead of the overall volume, we take the 
K\"ahler potential (\ref{Ktorus}) and expand $V_3$ in the region of the K\"ahler
cone where $\tau_1=\tau_2 = \tau_b \gg \tau_3 = \tau_s$
(as we explained above, at least two of the K\"ahler moduli have to become large 
simultaneously, if one wants to avoid any of the 2-cycle volumes 
becoming very small). 
This leads to (for simplicity setting
all $U_i=U$, all $\Et_i=\Et$ and all $\Es_i=\Es$):
\be\label{due}
V_3 &=& \frac{|W_0|^2 e^{K_{\mathrm{cs}}} }{2 S_1 \V^2}\Bigg\lbrace \left[ 
\frac{(\Et)^2+\tfrac{1}{2} (\partial_{U \bar U} K_{\rm cs})^{-1}
\partial_U \Et \partial_{\bar{U}} \Et}{8\tau_s^2 S_1^2}
+\mathcal{O}\left( \frac{1}{\tau_s^3} \right) \right] \nonumber \\[0.3cm]
&&+\frac{1}{\tau_b}\left[\frac{3 \,\tilde \xi \,S_1^{3/2}}{4\sqrt{\tau_s}}
+\frac{\Es}{\tau_s}+\frac{(\Et)^2+ (\partial_{U \bar U} K_{\rm cs})^{-1}
\partial_U \Et \partial_{\bar{U}}\Et } {4S^2_1\tau_s} +\mathcal{O}\left( 
\frac{1}{\tau_s^{3/2}}\right) \right]\nonumber\\
&& +\mathcal{O}\left(\frac{1}{\tau_b^2} \right)\Bigg\rbrace\ .
\ee
Obviously, the leading term in the large $\tau_b$ expansion now comes from the 
loop correction and not from the $\alpha'$ term (which term  
really dominates depends on the values of $S_1$ and $U$ as well, of course). 
Thus, an expansion of the potential as in LVS, cf.\ (\ref{largeVpot}), 
would not be realized in this case, even if one
found a way to lift enough zero modes by fluxes for $\tau_s$ to 
appear in a nonperturbative superpotential. 

This toy example was meant to show that  for 
a consistent large volume expansion in models with large
hierarchies in the K\"ahler moduli, it is important
to make sure that there are no 
correction terms in the K\"ahler potential (from loop or $\alpha'$-corrections) 
that are suppressed only by some of the small K\"ahler moduli. We should 
stress again that also terms  suppressed by the large volume can be 
dangerous if the suppression is less than for the $\alpha'$ term, i.e.\ 
if it is $\tau_b^{-\lambda}$ with $\lambda < 3/2$. 
The only exception to this rule is the case $\lambda=1$ as we showed above
(and as is shown more generally in appendices \ref{canc} and \ref{cancnp}). 
In this respect it would be important to know if the conjecture 
(\ref{realistic k}) really bears out. If it turns out that 
the actual form of the 1-loop corrections also contains 
terms like 
\be
\Delta K_{g_{\rm s}} \stackrel{?}{=} 
 \frac{t_b^{\lambda_1} t_i^{\lambda_2} \Et_i }{S_1 \cv}\ ,
\ee
with $\lambda_1 + \lambda_2 =1$ but $0<\lambda_1 \neq 1$ or $0$, 
such a 1-loop correction would spoil the large volume expansion 
performed in (\ref{largeVpot}).\footnote{In principle, one would also need 
an argument that no such terms arise at higher loop order, which 
would, however, have to be further suppressed in the dilaton $S_1$.} 


\subsection{Fibered Calabi-Yau manifolds}

The feature of orientifolds that all K\"ahler moduli 
appear linearly in the cubic expression for the volume
shows that a similar simplification can occur 
in the case of ($K3$) fibered Calabi-Yau manifolds, which also have the 
property that one K\"ahler modulus (the one corresponding to the 
volume of the base) only appears linearly
in the cubic expression for the volume. This takes the form
\be
\V = t_b \eta_{ij} t_i t_j + d_{ijk} t_i t_j t_k\ ,
\ee
where $\eta_{ij}$ are the intersection numbers of the ($K3$) fiber, and neither 
they nor the triple intersection numbers $d_{ijk}$ contain the 
index $b$, which is chosen to denote ``base'', but it is also 
suggestively the same index as the one we used for the large K\"ahler 
modulus in the $\mathbb{P}_{[1,1,1,6,9]}^4$ model.
Two-parameter examples of this type appear in e.g.\ 
\cite{Candelas:1993dm,Candelas:1994hw}.
In a region of the K\"ahler moduli space where the 
base $t_b$ is rather large but all the other $t_i$ stay relatively small, the 
volume is approximately $\V = t_b \eta_{ij} t_i t_j$. Thus, if the K\"ahler 
potential has a 1-loop correction $\sim \Et_b t_b/\V$, it 
could be approximated in this region by
\be \label{Eb}
\frac{\Et_b t_b}{\V} \quad \sim\quad \frac{\Et_b}{\tau_f} + 
\co \left(t_b^{-1}\right)\ ,
\ee
where $\tau_f = \eta_{ij} t_i t_j$ is the volume of the ($K3$) fiber
(which is small compared to $t_b^2$). Obviously, this would lead to 
a correction to the K\"ahler potential that is only suppressed by a 
single (small) 4-cycle volume, similar to the toroidal orientifold 
example we discussed in the last section. 

We should note that this limit (large base and small fiber for 
($K3$) fibered Calabi-Yau manifolds),
is quite different from
the one performed in the usual LVS of
\cite{Balasubramanian:2005zx}, even though
both cases involve hierarchical limits of the K\"ahler moduli. 
As explained in section \ref{sec:swiss}, the LVS limit consists in 
scaling the overall volume of the Calabi-Yau more or less 
isotropically while keeping holes in the 
bulk of the manifold small. 
In contrast, the limit of large base and small fiber is anisotropic. 
At the moment we have nothing to add about whether 
such anisotropic configurations with all moduli stabilized actually exist. We merely
wanted to point out that if they do exist, that would be an example
of smooth Calabi-Yau compactifications where
the $g_{\rm s}$ corrections we consider
dominate over the $\alpha'$ corrections considered in the large volume limit, 
as in the toroidal orientifold case.


\section{Further corrections}\label{sec:further}

In \cite{Conlon:2005ki}, further $\alpha'$ corrections 
to the string effective action 
beyond the one in \eqref{LVS}
were
considered. In the case of bulk $\alpha'$ corrections (i.e.\ those 
already present in type IIB bulk theory without D-branes, arising from 
sphere level) scaling arguments were given as to why they are 
suppressed in the large volume limit. 
Although that discussion was surprisingly
powerful in its simplicity, 
we do not consider it completely conclusive, if large hierarchies 
between the K\"ahler moduli exist. After all, 
dimensional analysis alone does not guarantee that 
the other $\alpha'$-corrections are always suppressed by additional 
powers in the overall volume, instead of powers of some of the 
small K\"ahler moduli. 
Moreover, in addition to the bulk $\alpha'$ corrections that appear at 
order ${\cal O} (\alpha'^3)$, in the models of interest for LVS 
further $\alpha'$-corrections arise on the worldvolume of D-branes and 
O-planes, cf.\ 
\cite{Green:1996dd,Dasgupta:1997cd,Cheung:1997az,Minasian:1997mm,Morales:1998ux,Stefanski:1998yx,Bachas:1999um,Fotopoulos:2001pt}.
These corrections begin already at order ${\cal O} (\alpha')$ and scaling
arguments of the kind used for the bulk corrections do not seem to be 
sufficient to neglect them. 

Indeed, there are correction terms involving two powers of the Riemann tensor
which do modify the effective D3-brane charge and tension, if the D7-branes 
are wrapped over 4-cycles with non-vanishing Euler number. These terms were already 
taken into account in \cite{Giddings:2001yu}. However, there are
further contributions to the DBI action at the same order in $\alpha'$, 
like $F_3^2 R$ or $F_3^4$, where $F_3$ stands for the RR 3-form
field strength, $R$ for the 
Riemann tensor and we left index contractions unspecified. If the D7-branes
do not break supersymmetry and remain BPS, it seems unlikely that these terms could 
contribute to the potential for the closed string moduli, 
i.e.\ induce some effective D3-brane tension. The reason 
is that there does not seem to be a corresponding term in the Chern-Simons action 
that could lead to the necessary modification of the effective D3-brane 
charge at the same time. This could be checked in more detail.

In general, we think that the question of additional corrections to the 
moduli (K\"ahler) potential deserves further attention.
Here we only outlined some steps in that direction.


\section{Conclusions}
In this paper, we have investigated whether string loop corrections
may impact {\it a)} stabilization in the large volume compactification
scenario (LVS), and {\it b)} the phenomenology of those scenarios,
as manifested in the soft supersymmetry breaking terms. The result is
that for the specific class of compactification manifolds considered
in LVS, so-called ``Swiss cheese'' Calabi-Yau manifolds, changes are
minuscule. Only if the loop corrections become abnormally large (in
the toroidal orientifold case, this can happen if the complex
structure is stabilized very large) do they affect LVS. 
For other classes of manifolds, the corrections may be important.
We 
hasten to add that the detailed expressions for the loop corrections
in LVS
remain unknown; we have merely tried to infer their scaling with the
K\"ahler moduli from experience in the toroidal orientifold
limit. We think it is important to attempt to address
this issue, as
the string coupling is stabilized at a nonzero value, so the
corrections cannot be turned off. 

We also stress the (to some readers obvious) fact that there remain a
host of issues that must ultimately be dealt with if one wishes
 to claim that these are ``string compactifications''. 
\begin{itemize}
\item
We cannot be sure that fluxes do not alter the corrections, since
backgrounds with 
RR and NSNS fluxes are not well understood in string perturbation
theory. 
\item
Additional corrections may appear (see section \ref{sec:further}) that
could be equally threatening to LVS as the loop corrections, or worse.
\item
In \cite{Berg:2005ja} only the corrections to the K\"ahler potential
coming from ${\cal N}=2$ sectors were determined and we based our 
generalization on those results. However, there might be 
interesting corrections coming from the ${\cal N}=1$ sectors as well.
\item
It has not yet been shown that a local Standard Model-like
construction can be embedded in the simplest examples like the
$\mathbb{P}_{[1,1,1,6,9]}^4$ model. If more general models turn out 
to be needed, one needs to reconsider
whether the requisite nonperturbative superpotentials are generated.
\item
We have largely ignored open string moduli, under the proviso that
they are stabilized heavy, as are the dilaton and complex structure
moduli.
\item
The coefficient $A(S,U)$ in the nonperturbative superpotential is
generally assumed to be of order 1. It is not known how generic this is.
\item
All string computations we have discussed were performed in a
supersymmetric context. In LVS supersymmetry is broken already before
uplift,
in the
AdS minimum. Supersymmetry breaking directly in string theory is not very well
understood \cite{Angelantonj:2002ct,Dudas:2004nd}. 
\end{itemize}
Faced with all these caveats, a pessimist might be inclined to give up. We
think we have shown that it is worth considering these issues in
detail. Sometimes, an effect one would have thought to be devastating
turns out to be as gentle as a summer breeze.


\section*{Acknowledgments}

It is a pleasure to thank Carlo Angelantonj, 
Vijay Balasubramanian, Massimo Bianchi, Joe Conlon, 
Gottfried Curio, Robbert Dijkgraaf, Michael Douglas, 
Bogdan Florea, Elias Kiritsis, Max Kreuzer, Fernando Marchesano, Peter Mayr, 
Thomas Mohaupt, Hans-Peter Nilles,  Gabi Pfuff, Fernando Quevedo,
Waldemar Schulgin, Mike Schulz, 
Stephan Stie\-ber\-ger, Angel Uranga, and Alexander Westphal for helpful 
discussions and comments and Boris K\"ors for initial collaboration. 
This work is supported in part by the European Community's Human 
Potential Program under 
contract MRTN-CT-2004-005104 'Constituents, fundamental forces and symmetries of the 
universe'. 
The work of M.~B.\ is supported by European Community's Human Potential
Program under contract MRTN-CT-2004-512194, `The European Superstring
Theory Network'. 
He would like to thank the 
Galileo Galilei Institute in Florence for hospitality. M.\ H.\ would 
like to thank the university of Nis for hospitality. 
The work of M.\ H.\ and E.\ P.\ 
is supported by the German Research Foundation (DFG) within 
the Emmy Noether-Program (grant number: HA 3448/3-1). 
Both M.~B.\ and M.~H.\ would like to thank the KITP
in Santa Barbara for hospitality during the 
program ``String Phenomenology'' and the university 
of Hamburg for hospitality during the 
workshop ``Generalized Geometry and Flux Compactifications''.


\appendix 

\section{Some details on LVS}

In this appendix we collect some details on the minimization of the
potential in LVS, mainly reviewing the results of 
\cite{Balasubramanian:2005zx,Conlon:2006tq}, but filling in some details. 
The minimization with respect to the axions (i.e.\
the imaginary parts of the K\"ahler moduli) is performed for an arbitrary 
number of K\"ahler moduli, while for the minimization of the 
real parts, we restrict to the example 
of the hypersurface in $\mathbb{P}_{[1,1,1,6,9]}^4$ discussed throughout the 
main text.


\subsection{LVS for $\mathbb{P}_{[1,1,1,6,9]}^4$}
\label{sec:11169}

Here we give some more numerical details on 
large-volume stabilization in the
$\mathbb{P}_{[1,1,1,6,9]}^4$ orientifold. The relevant features of this Calabi-Yau 
have been described in chapter \ref{sec:LVS}. The leading terms of the scalar potential are
\be\label{potte}
V \, e^{-K_{\rm  cs}} = \frac{\lambda \sqrt{\tau} e^{-2 a \tau}}{\V}
- \frac{\mu}{\V^2} \tau e^{-a \tau} + \frac{\nu}{\V^3}\ ,
\ee
where we use $\tau=\tau_s$ and $\V$ as the independent variables 
and for the expansion we have in mind the limit (\ref{taulnv}).
The minimum of this potential under the assumption that $a\tau\gg 1$ 
is given by
\be\label{sol}
\tau & = & \left( \frac{4 \nu \lambda}{\mu^2} \right)^{2/3},
\nonumber \\
\V & = & \frac{\mu}{2 \lambda} \left( \frac{4 \nu \lambda}{\mu^2}
\right)^{1/3}
e^{a \tau}\ .
\ee
In the $\mathbb{P}_{[1,1,1,6,9]}^4$ orientifold the coefficients $\lambda$, $\mu$ and 
$\nu$ can be calculated explicitly, yielding
\be
\lambda=\frac{12\sqrt{2}a^2 |A|^2}{S_1}\ , \,\,\,\,\mu=\frac{2a|AW_0|}{S_1} \,\,\,
\mathrm{and}\,\,\,\, \nu=\xi \frac{3}{8}\sqrt{S_1}|W_0|^2\ .
\ee
We notice that the value of $\tau$ at the minimum is determined only by the 
Euler number $\tau\propto \chi^{2/3}$ and the value of the dilaton 
$S_1$ at its minimum. An example of a set of possible parameters
(using $a=2\pi/10$, $A=1$, $S_1=10$ and $W_0=10$)
is
\be\label{param}
\xi=-\frac{\zeta(3)\chi}{2(2\pi)^3}\simeq 1.31\,\,\, &\longrightarrow&\,\,\, 
\nu\simeq 155\ ,\nonumber\\
\lambda\simeq 0.67\ ,\,&&\,\,\mu=\frac{4\pi}{10}\ .
\ee
There is an unknown overall 
factor $e^{K_{cs}}$ that does not change the shape of the potential 
and so leaves the position of the minima unchanged. 
For the parameters given in equation (\ref{param}),
 the minimum is at $\tau\simeq 41.1$ and $\V\simeq 9.96\cdot10^{11}$. These values 
come from equation (\ref{sol}) which is just approximated using the assumption 
$a\tau\gg1$.
This solution has the shortcoming that, if one is interested in the value of the 
potential at the minimum, after substitution of (\ref{sol}) into (\ref{potte}), 
one finds $V=0$. If instead one solves  the exact equation for the 
minimum of the potential numerically, 
the result is $V\simeq-6.6\cdot 10^{-37}$ at the point 
$\tau\simeq 41.7$ and $\V\simeq 1.38\,10^{12}$. From this one checks that, apart 
from the shortcoming that $V=0$, the approximate solution gives the position of 
the minimum with a good precision.


\subsection{Many K\"ahler moduli}
\label{sec:many} 

The simple picture of $\mathbb{P}_{[1,1,1,6,9]}^4$, gets slightly more involved
in models with more than two K\"ahler moduli, 
but some general statements can still be made. 
For a single small K\"ahler modulus, among the leading contributions to the 
potential only the one from $V_{\rm np2}$ is axion dependent, 
while the leading terms in $V_3$ and $V_{\rm np1}$ are axion independent.
For several small K\"ahler moduli, 
all three terms are axion dependent. However, the
argument that the leading term in $V_{\rm np2}$ only receives a sign change
due to axion stabilization
generalizes (and holds also for the
regular KKLT scenario with relatively small volume, see e.g.
\cite{Conlon:2006tq}, section 3.2). 

Indeed, with the superpotential (\ref{supo}) one obtains
\be
V_{\rm np1}&=&e^{K}\, G^{\bar{\jmath} i}\left[a_ia_j|A_iA_j|e^{-a_i\tau_i-a_j\tau_j}
\cos (-a_ib_i+a_jb_j+\beta_i-\beta_j)\right]\ ,\nonumber \\[3ex]
V_{\rm np2}&=& -2e^{K}\, a_iG^{\bar{k} i} K_{\bar k} \Big[ |A_iW_0|
e^{-a_i\tau_i}\cos(-a_ib_i+\beta_i-\beta_{W_0})\nonumber \\
&& \hspace{2.4cm} + |A_iA_j|e^{-a_i\tau_i-a_j\tau_j}
\cos(-a_ib_i+a_jb_j+\beta_i-\beta_j)\Big]\ , \\[3ex]
V_{3}&=&e^{K}\, (G^{\bar{k} l}K_{\bar{k}}K_{l}-3)\, \Big[ |W_0|^2 + 2\,|W_0A_i|
e^{-a_i\tau_i}\cos(-a_i b_i+\beta_i-\beta_{W_0})\nonumber \\
&& \hspace{3.1cm} + |A_iA_j|e^{-a_i\tau_i-a_j\tau_j}\cos (-a_ib_i+a_jb_j+\beta_i-\beta_j)\Big]\ 
\nonumber , 
\ee
where $A_i=|A_i|e^{i\beta_i}$, $W_0=|W_0|e^{i \beta_{W_0}}$ and 
a sum over repeated indices is understood throughout.
As the only dependence on the axions is in form of cosines, one can easily 
see that this potential has a minimum for 
\be \label{axionmin}
a_ib_i=-\beta_{W_0}+\beta_i+ n_i \pi\ \quad , \quad n_i \in 2 \mathbb{Z} + 1\ .
\ee
We notice that 
the minimum of the $b_i$ depends on the (already fixed) complex structure 
moduli, but it is independent of the K\"ahler moduli.

In the regime (\ref{taulnv}) the scalar potential again contains three terms
at leading order,  
\be 
V_{\rm np1}&\sim& 2 e^{K_{\mathrm{cs}}}\,\frac{a_i a_j |A_iA_j| e^{-a_i \tau_i-a_j\tau_j}
M_{j}^l M_{i}^k (- \V \V_{lk} + \V_l \V_k)}{S_1 \V^2}+ \ldots\ , \nonumber \\
V_{\rm np2}&\sim&- 2 e^{K_{\mathrm{cs}}}\,\frac{a_i 
|A_i\,W_0| e^{-a_i\tau_i} \tau_i}{S_1 \V^2}+ \ldots \ ,  \label{Vmany} \\
V_3&\sim& e^{K_{\mathrm{cs}}}\,\frac{3\xi S_1^{1/2}}{8\V^3}|W_0|^2+ \ldots\ ,\nonumber 
\ee
where the sum over $i$ and $j$ effectively only picks up terms from the 
small moduli because of the exponential suppression of $V_{\rm np1}$ and 
$V_{\rm np2}$. Moreover, for $V_{\rm np1}$ we used the 
form (\ref{ginv}) for the inverse of the moduli metric with respect to the 
basis (\ref{taua}). The K\"ahler moduli appearing in the 
nonperturbative superpotential are linear combinations of these, which we account 
for by a basis-changing matrix $M_{i}^k$, i.e.
\be \label{transform}
T_i = M_i^k\, \tilde T_k\ ,
\ee
where $\tilde T_k$ are the fields defined in (\ref{taua}) and $T_i$ 
are the K\"ahler moduli appearing in the nonperturbative superpotential. (Another 
way of saying this is that the real parts of $T_i$
 measure the volumes of 
a basis of divisors that have the right properties to contribute to the 
nonperturbative superpotential.)  
In the second term we used 
\be \label{GK}
G^{\bar{k} i} K_{\bar k} = - 2 \tau_i + \ldots = -2 \,{\rm Re}\, T_i + \ldots \ .
\ee
In the basis (\ref{taua}), this would follow straightforwardly 
from (\ref{ki}), (\ref{ginv}) and the relations (\ref{Krels}), 
but it  holds equally well after a change of basis, because both sides of (\ref{GK})
transform linearly under a change of basis (\ref{transform}). 

Note finally that the ellipsis in (\ref{Vmany}) and (\ref{GK}) 
stand for subleading corrections in the large volume 
limit (assuming also (\ref{taulnv})).


\section{Loop corrected inverse K\"ahler metric for $\mathbb{P}_{[1,1,1,6,9]}^4$} 
\label{loopcorrected}

We now have a closer look at the inverse metric from the K\"ahler 
potential in equation (\ref{realistic k}). 
We invert the $4\times4$ matrix and focus on the four terms that appear 
in the scalar potential for the K\"ahler moduli,
\be\label{inv metric}
G^{\bar b b}&= &\frac{4}{3}\tau_b^2 + \mathcal{O}\left( \tau_b \right) \ ,\nonumber \\
G^{\bar b s} = G^{\bar s b} &= &4\tau_b\tau_s \left( 1 + \frac{6 \Et_s}{\Delta} \right)
+ \mathcal{O}\left( \tau_b^0\right) \ , \\
G^{\bar s s}&= &\frac{8}{3}\tau_b^{3/2}\sqrt{\tau_s}\,\frac{\sqrt{2} S_1 \tau_s}{\Delta}
+ \mathcal{O}\left(  \tau_b^{1/2}\right) \nonumber \ ,
\ee
where we have performed an expansion in $\tau_b\simeq\V^{2/3}$ and the quantity 
$\Delta$ was introduced in (\ref{Delta}).
We notice that only $G^{\bar b b}$
is not corrected at leading order. The apparent divergence from zeros 
of the denominator $\Delta$ is an artifact of 
the expansion. In fact, the determinant of the 
(entire) K\"ahler metric behaves as
\be 
{\rm det}\, G \sim A \tau_b^{-7/2} + B \tau_b^{-9/2} + \ldots
\ee
for some expressions\footnote{This $A$ has nothing to do with the $A$ in $W_{\rm np}$.} $A$ 
and $B$, which depend on the moduli $\tau_s, U$ and $S_1$. In particular, one finds
$A \sim \Delta$, but $B$ does not vanish at a zero of $\Delta$.  
Thus, in general the expansion in large $\tau_b$ picks up 
the factor $A$, which is responsible 
for the apparent divergence in \eqref{inv metric}. However, 
this is fictitious because when $\Delta=0$, the next term proportional to
$B$ is non-vanishing and the determinant stays away from zero. 
Indeed, we do not expect to find any zero 
of the determinant in the range of validity of the parameters.

If $\Et_s \ll (S_1 \tau_s)$, one can further expand (\ref{inv metric}) 
with respect to $\Et_s / (S_1 \tau_s)$, yielding 
\be
G^{\bar b b}&= &\frac{4}{3}\tau_b^2 +\mathcal{O}\left(  \tau_b\right)  \nonumber\,, \\
G^{\bar b s} = G^{\bar s b} &=& 4\tau_b\tau_s \left( 1 + \frac{3 \sqrt{2} \Et_s}{S_1 \tau_s}
+\mathcal{O}\Big( \frac{(\Et_s)^2}{S_1^2 \tau_s^2}\Big) \right) \nonumber\,, \\
G^{\bar s s}&=& \frac{8}{3}\tau_b^{3/2}\sqrt{\tau_s}
\left( 1 + \frac{3 \Et_s}{\sqrt{2} S_1 \tau_s}
+ \mathcal{O}\Big( \frac{(\Et_s)^2}{S_1^2 \tau_s^2}\Big) \right)\ . \nonumber \\
\ee
Depending on the values of the moduli ($\tau_s$ and $S_1$), this 
expansion may or may not be useful. In general, only the expansion 
in $\tau_b$ makes sense and one has to deal with the full expressions 
(\ref{inv metric}). That is what we did in section \ref{P model}.


\section{No-scale K\"ahler potential in type II string theory}\label{noscale} 

In this appendix we review why compactification of type IIA and type IIB theory on 
general Calabi-Yau manifolds, or orientifolds thereof, 
lead to no-scale (F-term) potentials  
if 
\be \label{assumpta}
\mbox{{\it i)} the superpotential does not depend on the K\"ahler moduli}
\ee
 and if 
\be \label{assumptb}
\mbox{{\it ii)} one uses the 
tree-level form of the K\"ahler potential.}
\ee
(Of course, in LVS neither $i)$ nor $ii)$ holds, but one
can think of jointly imposing $i)$ and $ii)$ as a zeroth-order approximation,
that we will successively move away from in later subsections
of this appendix.)

If  the moduli spaces of K\"ahler and complex structure moduli factorize
(see appendix \ref{calcu} for more details on this),
and under assumption {\it i)}, the F-term potential takes the form
\be\label{pot}
V & = & e^{K}\left( G^{ \bar I J} D_{\bar I}\bar WD_{J}W-3|W|^2 \right) \\ 
& = & e^{K} \left( G^{ \bar a b} D_{\bar a}\bar WD_{b}W +(G^{\bar{\imath} j} 
K_{\bar{\imath}} K_{j} - 3)|W|^2 \right)\ .
\ee
The indices $a$ and $b$ run over the complex structure moduli and the dilaton, $i,j$ over the 
K\"ahler moduli and $I$ and $J$ refer to all moduli. 


The condition for a no-scale potential ($V=0$ for the K\"ahler
moduli) is then
\be \label{noscale2}
G^{\bar{\imath} j}K_{\bar{\imath}}K_{j}=3\ ,
\ee
and we will verify in turn that this is fulfilled in both 
type IIA and type IIB Calabi-Yau compactifications, 
if one uses the tree-level K\"ahler potential, as in assumption {\it
  ii)}. In that case, the moduli spaces of K\"ahler and complex
structure
moduli do factorize exactly.


\subsection{No-scale structure in type IIA}
\label{noscale2a}

The tree level K\"ahler potential for the K\"ahler moduli is
\be\label{ka}
K=-\ln \left[\frac{1}{48}d_{ijk}(\sigma+\bar{\sigma})_i(\sigma+\bar{\sigma})_j
(\sigma+\bar{\sigma})_k\right]=-\ln \left[\frac{1}{6}d_{ijk}t_it_jt_k\right] = -\ln(\V) \ ,
\ee
where $d_{ijk}$ are the intersection numbers of the Calabi-Yau, 
\be
d_{ijk}=\int_{CY} \omega_i \wedge \omega_j \wedge \omega_k\ ,
\ee
and 
\be
\sigma_i = t_i + i  c_i
\ee
are the complexified K\"ahler moduli whose real parts $t_i$ represent the volumes of 
2-cycles and whose imaginary parts originate from the expansion of the
NSNS 2-form. Using the K\"ahler form 
\be
J= t_i \omega_i
\ee
of the Calabi-Yau, it is useful to introduce the notation
\be
\V & = & \frac16 \int J \wedge J \wedge J= \frac16 d_{ijk} t_i t_j t_k\ , \non
\V_i & = & \frac{1}{2} \int \omega_i \wedge J \wedge J= \frac{1}{2} d_{ijk} t_j t_k \ ,\non
\V_{ij} & = & \int \omega_i \wedge \omega_j\wedge J= d_{ijk} t_k\ .
\ee
Note that here the index $i$ does not 
denote a derivative with respect to 
the K\"ahler variables (in contrast to subscripts on 
the K\"ahler potential $K$). Instead, one has the relations
$\V_i= 2 \partial_{\sigma_i} \V$ and 
$\V_{ij}=4 \partial_{\sigma_i}\partial_{\bar \sigma_{\bar{\jmath}}} \V$. 
It is straightforward to calculate 
\be\label{K}
K_{i}=-\frac{\V_i}{2\V}=K_{\bar{\imath}} \quad G_{i\bar{\jmath}}=K_{i \bar{\jmath}}= 
- \frac14 \left( \frac{\V_{ij}}{\V} - \frac{\V_i\V_{j}}{\V^2} \right)\ .
\ee
Then one can show that the inverse K\"ahler metric is
\be
G^{\bar{\jmath} i}=-4 \V^{ji}\V+ 2 t_j t_i\ .
\ee
To verify this, one has to use 
\be \label{Krels}
\V^{ij}\V_{j}=\frac12 t_i\ , \quad \V_{ij} t_j= 2 \V_i\ , \quad  \V_i t_i= 3 \V\ .
\ee
Putting everything together, one arrives at
\be
G^{\bar  \imath j} K_{\bar{\imath}} K_{j} =\left[ -4 \V^{ij}\V+ 2 t_i t_j \right] \frac14 \frac{\V_i}{\V} \frac{\V_{j}}{\V} =3\ ,
\ee
i.e.\ (\ref{noscale2}) is fulfilled
under assumption \eqref{assumptb}.  


\subsection{No-scale structure in type IIB}
\label{ns2b}

In the type IIB case, the tree-level K\"ahler potential for the K\"ahler moduli is
\be \label{K2a}
K=-2\ln \left[ \frac{1}{6}d_{ijk}t_i t_j t_k \right]=-2\ln (\V )\ .
\ee
The difference to the IIA case is that, even if $K$ 
in (\ref{K2a}) is expressed in terms of the 2-cycle volumes $t_i$, the real parts of the good K\"ahler moduli, $\tilde T_i$, are now the 4-cycle 
volumes $\tilde \tau_i$ (the imaginary parts, on the 
other hand, arise from the RR 4-form). The relation between them 
depends on the particular Calabi-Yau:
\be \label{taua}
{\rm Re}\, \tilde T_i = \tilde \tau_i =\frac12 d_{ijk} t_j t_k= \V_i\ ,
\ee
which cannot be inverted in general.\footnote{Note that the 
K\"ahler moduli appearing in the 
non-perturbative superpotentials in the examples of \cite{Denef:2004dm} are 
related to the ones in (\ref{taua}) by a linear field redefinition. 
However, this does not play any role in verifying the no-scale structure at 
leading order, as (\ref{noscale2b}) below is invariant under field 
redefinitions. We chose to make the distinction 
clear by using tildes for the K\"ahler moduli defined by (\ref{taua}).}
In order to calculate $K_{ i}=\partial_{\tilde T_i}K$ we note that 
\be \label{dva}
\partial_{t_i} &=& (\partial_{t_i} \tilde T_j) \partial_{\tilde T_j} 
+ (\partial_{t_i} \bar{\tilde T}_{\bar{\jmath}}) 
\partial_{\bar{\tilde T}_{\bar{\jmath}}} \non
&=& \V_{ij} \left(  \partial_{\tilde T_j} 
+ \partial_{\bar{\tilde T}_{\bar{\jmath}}} \right)\ .  
\ee
If acting on a function $F$ that only depends on $\tilde T 
+ \bar{\tilde T}$, as is the case for $K$,  
(\ref{dva}) simplifies to 
\be
\partial_{t_i} F(\tilde T + \bar{\tilde T}) = 2 \V_{ij} 
\partial_{\tilde T_j} F(\tilde T +\bar{\tilde T})\ ,
\ee
where on the left hand side $\tilde T$ is understood as a function of $t$. 
Alternatively, one has
\be
\partial_{\tilde T_i} F(\tilde T + \bar{\tilde T}) 
= \frac{1}{2} \V^{ij} \partial_{t_j} F(\tilde T + \bar{\tilde T})
=\partial_{\bar{\tilde T}_{\bar{\imath}}} F(\tilde T + \bar{\tilde T})\ .
\ee
Using this, one can calculate
\be \label{ki}
K_{i}=-\frac{2}{\V} \partial_{\tilde T_i} \V =  - \frac{ \V^{ij} \V_j}{\V}
=- \frac{ \,\,t_i}{2\,\V} = K_{\bar \imath} \ ,
\ee
where in the last step we used (\ref{Krels}). In the same way one can calculate
\be
G_{i \bar{\jmath}} = \frac{1}{4} \left( - \frac{\V^{ij}}{\V} 
+ \frac12 \frac{t_i t_j}{\V^2} \right)\ .
\ee
Using this formula one can check that the inverse K\"ahler metric is given by
\be \label{ginv}
G^{\bar{\imath}j}= 4 \left( - \V \V_{ij} + \V_i \V_j \right)\ .
\ee
Putting everything together, no-scale structure holds also for type IIB:
\be \label{noscale2b}
G^{\bar  \imath j} K_{\bar{\imath}} K_{j} = \left( - \V \V_{ij} 
+ \V_i \V_j \right) \frac{t_i}{\V} \frac{t_j}{\V} = 3\ ,
\ee
again under the assumption \eqref{assumptb}.


\subsection{Cancellation with just the volume modulus}  \label{canc ap}

Now we relax assumption \eqref{assumptb}.
For simplicity, let us first 
consider the K\"ahler potential 
\be
K=-3\, \mathrm{ln}(T+\bar{T})+ \frac{\Xi}{(T+\bar{T})^\lambda}\ ,
\ee
which corresponds to the case of a single K\"ahler modulus and the complex 
structure moduli and the dilaton
are neglected. A generic quantum correction was added to the tree level term, which   
could be an $\al$ or a loop correction, depending on the value of $\lambda$. 
Focusing on $V_3$, i.e.\
\be
\frac{V_3 }{e^K\,|W|^2}=G^{\bar{\jmath}i} K_{\bar{\jmath}} K_{i}-3\ ,
\ee
one calculates
\be\label{can1}
\frac{V_3 }{e^K\,|W|^2}&=&\frac{(3(2\tau)^{\lambda}+\xi \lambda)^2}{3(2\tau)^{2\lambda}
+\Xi(2\tau)^{\lambda} \lambda(\lambda+1)}-3\nonumber \\ &=& 3 - 3 
+\frac{(\lambda-\lambda^2)\Xi}{(2\tau)^\lambda}+\frac{\Xi^2\lambda^4}{3(2\tau)^{2\lambda}}
+\mathcal{O}\left( \frac{1}{\tau^{3\lambda}}\right).
\ee
This simplified calculation gives an intuition of why the $\Et_b$-term 
does not appear in $V_3$ of \eqref{Vwnpzero} 
whereas the $\al$- and $\Et_s$-terms contribute. When the exponent of the 
quantum correction is exactly $1$, there is a cancellation at leading order 
in the scalar potential (compare also the discussion in footnote \ref{cancelfn}). 
Note that since we focused on $V_3$ in this subsection,
it did not matter whether assumption \eqref{assumpta} holds or not.


\subsection{Cancellation with many K\"ahler moduli}\label{canc}

We would now like to see how the previous result is changed when we have 
an arbitrary number of moduli. We 
do not make \textit{any} assumption on the dependence of the volume on the 
K\"ahler moduli (``Swiss cheese'' or fibered manifolds are special cases). 
Due to its relevance for LVS, we consider a single correction to the 
K\"ahler potential which only depends on the large K\"ahler modulus $T_b$
(an example would be the $\al$-correction or the loop term proportional 
to $\Et_b$, considering the moduli other than the K\"ahler moduli 
as fixed; this is 
allowed at leading order in a $\tau_b$-expansion, as we argue in 
appendix \ref{calcu}). Thus, we take the K\"ahler potential 
to be of the form
\be\label{kal}
K=K^{(0)} + \delta K = -2\mathrm{ln}(\V)+\delta K(T_b,\bar T_b)\equiv-2\mathrm{ln}(\V)
+ \frac{\Xi_b}{(T_b+\bar{T}_b) ^{\lambda}}\ .
\ee
Again focusing on $V_3$, we obtain
\be
\frac{V_3 }{e^K |W|^2}&=&G^{\bar \jmath i} K_{\bar \jmath} K_{i}-3=(G^{ \bar \jmath i}_0
+\delta G^{ \bar \jmath i} )
\left[K^{(0)}_{\bar \jmath} + \delta K_{\bar \jmath}\right]
\left[K^{(0)}_i+\delta K_i \right]-3\nonumber \ , \\
\ee
where $\delta K_{i}\equiv \partial_{T_i} \delta K$ and $G_0^{\bar \jmath i}$ 
is the inverse metric of appendix 
\ref{ns2b}; finally $\delta G^{ \bar \jmath i}$ is the modification of the 
inverse metric coming from considering the modified K\"ahler potential \eqref{kal}. 
Explicitly one has
\be  \label{inversemetric}
G^{\bar \jmath i}&=&(G^0_{i \bar \jmath }+\delta K_{i \bar \jmath})^{-1}\simeq 
G^{\bar \jmath i}_0-G^{\bar \jmath h}_0 
\delta K_{h \bar k} G^{\bar{k} i}_0 +\ldots\ , \nonumber \\
\delta K_{i}&=&-\frac{\lambda \,\Xi_b}{(2\tau_b)^{\lambda+1}}\delta_{ib}\ ,\qquad 
\qquad \delta K_{i \bar \jmath}
= 
\frac{(\lambda^2+\lambda)\,\Xi_b}{(2\tau_b)^{\lambda+2}}\delta_{ib}\delta_{\bar{\jmath}b}\ .
\ee
We now put everything together and use the results of appendix \ref{ns2b} and 
formula (\ref{GK}) (which, for the unperturbed metric and K\"ahler potential, 
is an exact equality) to arrive at
\be
\frac{V_3 }{e^K |W|^2}&=&\left[G_0^{\bar \jmath i} K^{(0)}_{\bar \jmath}
K^{(0)}_i -3\right]
+2\left[ G_0^{\bar \jmath i} K^{(0)}_{\bar \jmath} \delta K_{i} 
\right]-\left( G_0^{\bar \jmath k}\delta K_{k \bar h} G_0^{\bar {h} i}\right)
K^{(0)}_{\bar \jmath} K^{(0)}_i 
+\dots \nonumber\\[.5cm]
&=&0+\frac{4\lambda \Xi_b}{(2\tau_b)^{\lambda+1}}\tau_b-
\frac{4(\lambda^2+\lambda)\Xi_b}{(2\tau_b)^{\lambda+2}}\tau_b \tau_b + \dots \nonumber \\
&=& \frac{(\lambda-\lambda^2)\Xi_b}{(2\tau_b)^{\lambda}}+\dots\ .
\ee
We notice that the term $1/\tau_b^\lambda$ vanishes exactly for 
$\lambda=1$, independently of the explicit form of the volume in terms of the 
K\"ahler moduli. In particular, the loop correction proportional to $\Et_b$ experiences a 
cancellation at leading order in $V_3$ (and it is not difficult
to see that the subleading order 
is suppressed by $\tau_b^{-2 \lambda}$).
Therefore, the loop correction is subleading in the potential 
compared to the $\al$ correction, 
even though it is leading in the K\"ahler potential. Next, 
we would like to extend this analysis to the other parts of the 
potential, i.e.\ $V_{\rm np1}$ and $V_{\rm np2}$.

\subsection{Perturbative corrections to $V_{\rm np1}$ and $V_{\rm np2}$}
\label{cancnp}

We now introduce the nonpertubative superpotential into the game,
i.e.\ relax assumption \eqref{assumpta},
 and look 
at the other terms of the scalar potential, $V_{\rm np1}$ and  $V_{\rm
  np2}$ (see eq.\ \eqref{alcuni}). 
For this, we restrict to the $\mathbb{P}_{[1,1,1,6,9]}^4$ model again. 
The contribution $V_{\rm np1}$ is proportional to $G^{\bar s s} \bar{W}_{,\bar s} W_{,s}$. 
From \eqref{inv metric} we see that no $\Et_b$ appears at leading order. This 
can be understood as follows. Consider the K\"ahler potential 
\eqref{kal} where now $\V$ is the volume of $\mathbb{P}_{[1,1,1,6,9]}^4$, 
given in \eqref{tbts}. Then the scaling with the large K\"ahler modulus $\tau_b$ 
is schematically given by
\be 
G^{ \bar s s}&\simeq &G^{ \bar s s}_0 - G^{\bar s b}_0 
\delta K_{b \bar b}G^{\bar b  s}_0+ \ldots \nonumber \\
&\sim&\tau_b^{3/2}+\tau_b^2\frac{\Xi_b}{\tau_b^{\lambda+2}}
\sim \tau_b^{3/2}+\frac{\Xi_b}{\tau_b^{\lambda}}+ \ldots \ ,
\ee
which shows that any loop correction to the K\"ahler potential
of the form $\Xi/\tau_b^{\lambda}$ leads to a 
subleading contribution to $V_{\rm np1}$
in the large volume expansion. As usual, the ellipsis stands for 
terms that are even more subleading  in the $\tau_b$ expansion. 

To understand the $\Et_s$ correction to $G^{\bar s s}$ we need to consider
\be\label{kal tilde}
K=-2\ln (\V)+\tilde{\delta}K(T,\bar T)\equiv-2\ln(\V)
+ \frac{\Xi_b\, g(T_s,\bar{T}_s)}{(T_b+\bar{T}_b) ^{\lambda} }
\ee
for some function $g(T_s,\bar{T}_s)$ of the small K\"ahler modulus
and we assume $\lambda \geq 3/2$ in the following. 
Then, again very schematically, the scaling behavior is given by\footnote{For 
$\lambda=3/2$ we can still use the expansion of the inverse metric \eqref{inversemetric}, 
because the correction term would also be further 
suppressed e.g.\ in the dilaton.}
\be 
G^{ \bar s s}&\simeq &G^{ \bar s s}_0-G^{\bar s i}_0 
\tilde{\delta}K_{i \bar \jmath}G^{\bar \jmath  s}_0 + \ldots \nonumber \\
&\sim&\tau_b^{3/2}+\Xi_b\,(\tau_b,\tau_b^{3/2}) 
        \left(\begin{array}{cc} \tau_b^{-\lambda-2}&\tau_b^{-\lambda-1}\\
                                \tau_b^{-\lambda-1}&\tau_b^{-\lambda}
              \end{array}
              \right) 
              \left(
              \begin{array}{c} \tau_b\\
                                \tau_b^{3/2}
              \end{array} 
              \right) \label{np pert} + \ldots \\
&\sim&\tau_b^{3/2}+\tau_b^{-\lambda}+\tau_b^{-\lambda+3/2}+\tau_b^{-\lambda+3}
+ \ldots \ . \nonumber
\ee
One sees that $\lambda=3/2$ indeed contributes at the same order as $G^{\bar ss}_0$. 
This is confirmed by the dependence of $G^{\bar ss}$ in \eqref{inv metric} 
on $\Et_s$ through $\Delta$, cf.\ \eqref{Delta}.

We now consider $V_{\rm np2}$. 
This is proportional to $G^{\bar \jmath s} K_{\bar \jmath}$. 
Again we start by considering a correction to the K\"ahler potential 
whose only dependence 
on the K\"ahler moduli is via $\tau_b$, as in \eqref{kal}. Schematically, we find
\be 
G^{\bar \jmath s} K_{\bar \jmath}&\simeq&(G^{\bar \jmath s}_0
-G^{\bar \jmath b}_0 \delta K_{b \bar b} G^{\bar b s}_0)
\left[K^{(0)}_{\bar \jmath} + \delta K_{\bar \jmath}\right]+ \ldots \nonumber\\
&\sim& \tau_s+\frac{G^{\bar s b}_0\,\Xi_b}{\tau_b^{\lambda+1}} + \ldots
\sim \tau_b^0+\frac{\Xi_b}{\tau_b^{\lambda}}+ \ldots \ .
\ee
This result is confirmed by the absence of $\Et_b$ in the leading term of 
$V_{\rm np2}$. A calculation very similar to the one in \eqref{np pert} 
shows, however, that $V_{\rm np2}$ {\it is} modified by a 
correction to the K\"ahler potential of the form \eqref{kal tilde} for 
$\lambda=3/2$. It is straightforward to 
generalize this analysis to a more general form of the ``Swiss cheese'' volume, 
with more than one small K\"ahler modulus.


\section{KK spectrum with fluxes}
\label{E7neq0}

In this section we would like to develop some intuition on how the analysis 
of sections \ref{P model} and \ref{sec:gaugino} might change in the 
presence of fluxes. We will restrict the discussion to one 
possible effect of the fluxes, namely
their influence on the KK spectrum. It is not known explicitly how 
closed string fluxes, which are present in LVS, would change the mass 
spectrum. We will consider a toy example, using an  analogy to 
the correction arising from world volume fluxes (cf.\ \cite{Blumenhagen:2006ci}), 
in order to get a feeling for what kind
of effects one might expect. In particular, 
for the purposes of this appendix we assume 
a modified KK mass spectrum 
of the form
\be \label{KKcorr}
m_{\rm KK}^2 \sim \frac{1}{t_{\rm str} \left(1 + \frac{F^2}{t_{\rm str}^2}\right)} = 
\frac{\sqrt{S_1}}{t \left(1 + \frac{F^2 S_1}{t^2} \right)} \ ,
\ee
where $F$ represents any of the fluxes that may be present, and in the second 
equality the factors of $S_1$ appeared when expressing
the 2-cycle volumes in Einstein frame as compared to the string 
frame ($t \sim e^{-\Phi/2} t_{\rm str}$). Note that expanding 
\eqref{KKcorr} for large values of $t$ would lead to a correction
$\Delta m^2 \sim F^2/t^3$, whose scaling with the flux and with $t$
is reminiscent of the moduli masses induced by closed string 3-form 
flux \cite{Kaloper:1999yr,Kachru:2002he}. In that case,
the suppression would be by the overall volume (which would lead to 
only mild effects in LVS), but in \eqref{KKcorr}
we allow for a suppression by single 2-cycle volumes (which might be 
the small 2-cycle in the $\mathbb{P}_{[1,1,1,6,9]}^4$ model).

Substituting (\ref{KKcorr}) in \eqref{generalize},
we now consider the scalar potential resulting from
\be \label{realistic k2}
K&=&-\mathrm{ln}(2S_1)-2\,\mathrm{ln}(\V)+K_{\rm
  cs}(U,\bar U) - \frac{\tilde{\xi} S_1^{3/2}}{  \V}
  +\frac{\sqrt{\tau_b} \Et_b }{ S_1\V}
  + \frac{\sqrt{\tau_s} \Et_s }{ S_1\V} \left( 1 + \frac{F^2 S_1}{\tau_s} \right)\ , 
  \nonumber \\ [1mm]
W &=&  W_0 +  A e^{-a T_s}\ .
\ee 
We have not included any flux correction to the term 
proportional to $\Et_b$ because we
expect such corrections 
to be subleading in a large volume expansion.\footnote{Even though we think it is
unlikely, we cannot
exclude that the correction to KK masses that scale like 
$t_b^{-1}$ without fluxes is only suppressed 
by $F^2/\tau_s$ instead of $F^2/\tau_b$. In that 
case, one would have to redo the analysis of appendix \ref{cancnp}, 
using \eqref{kal tilde} with $\lambda=1$. This would prohibit the 
use of the expansion \eqref{inversemetric}, because in the large volume limit
the leading contribution to $G_{s\bar s}$ would arise from 
the loop correction (it would scale as $\tau_b^{-1}$ as opposed to the 
scaling of the tree level contribution $\sim \tau_b^{-3/2}$). In that case
the leading terms in $V_{\mathrm{np1}}$ and $V_{\mathrm{np2}}$ would be suppressed
compared to $V_3$ and only arise at order $\V^{-10/3}$, thus invalidating the 
volume expansion of LVS.} 
Note that the $F$-dependent correction term 
we did include
is of the same form as the
 winding string correction $\sim \Es_s$, when one neglects any potential 
flux corrections  to the winding string spectrum, cf.\ \eqref{LVSLoop}
(remember that $\gw^{s}$ would just be proportional to $1/\sqrt{\tau_s}$
without fluxes). 
Thus, by considering \eqref{realistic k2}
we implicitly also analyze in the following the effect of corrections from 
winding strings 
(recall from section \ref{P model} that this correction is not
present in the $\mathbb{P}_{[1,1,1,6,9]}^4$ model, but may be present
in general).

\begin{figure}[ht] \begin{center}
\psfrag{E}[tc][bc][1][1]{$E$}
\psfrag{S1}[tc][bc][1][1]{$S_1$}
\includegraphics[width=0.4\textwidth]{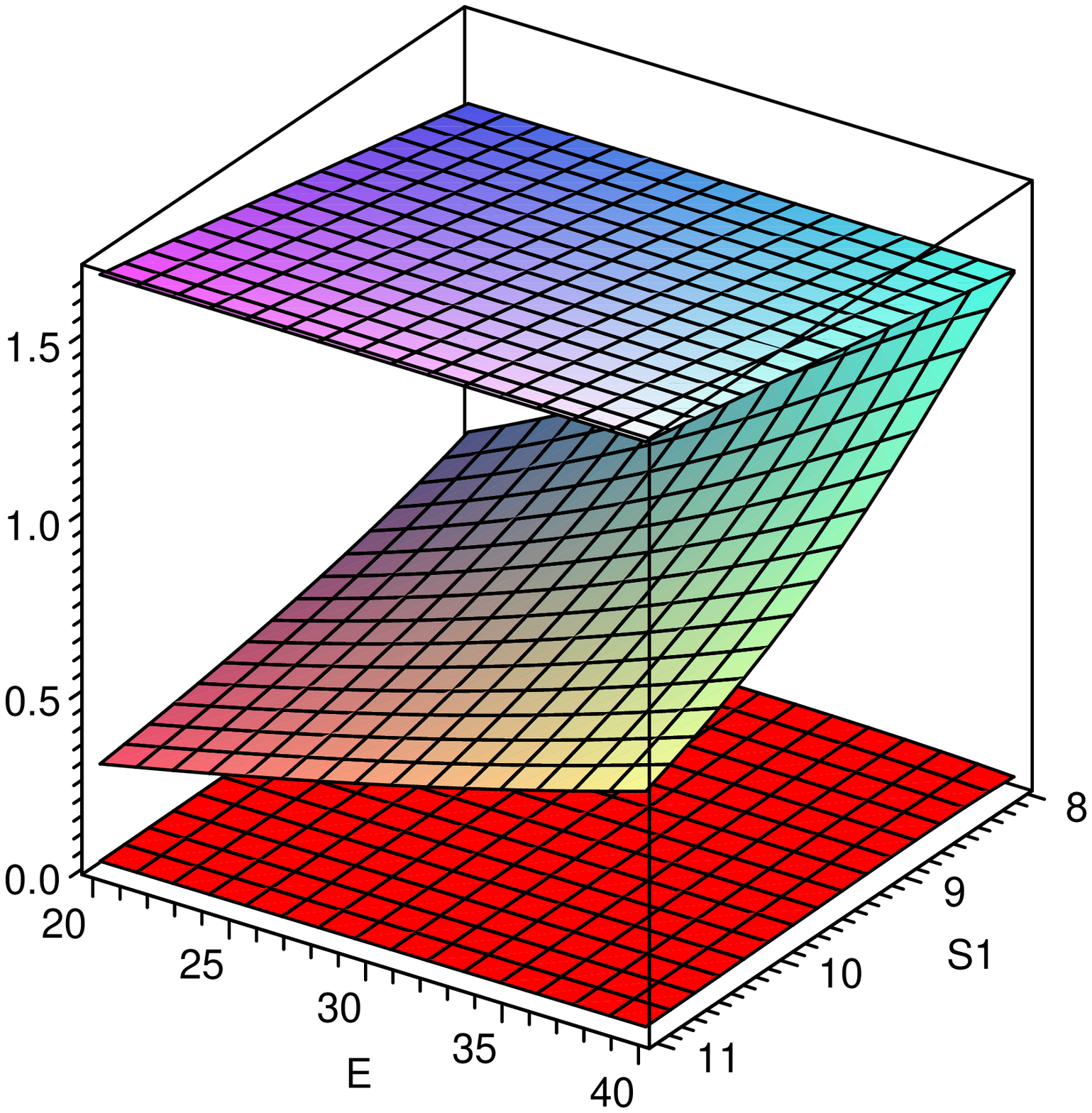}
\includegraphics[width=0.4\textwidth]{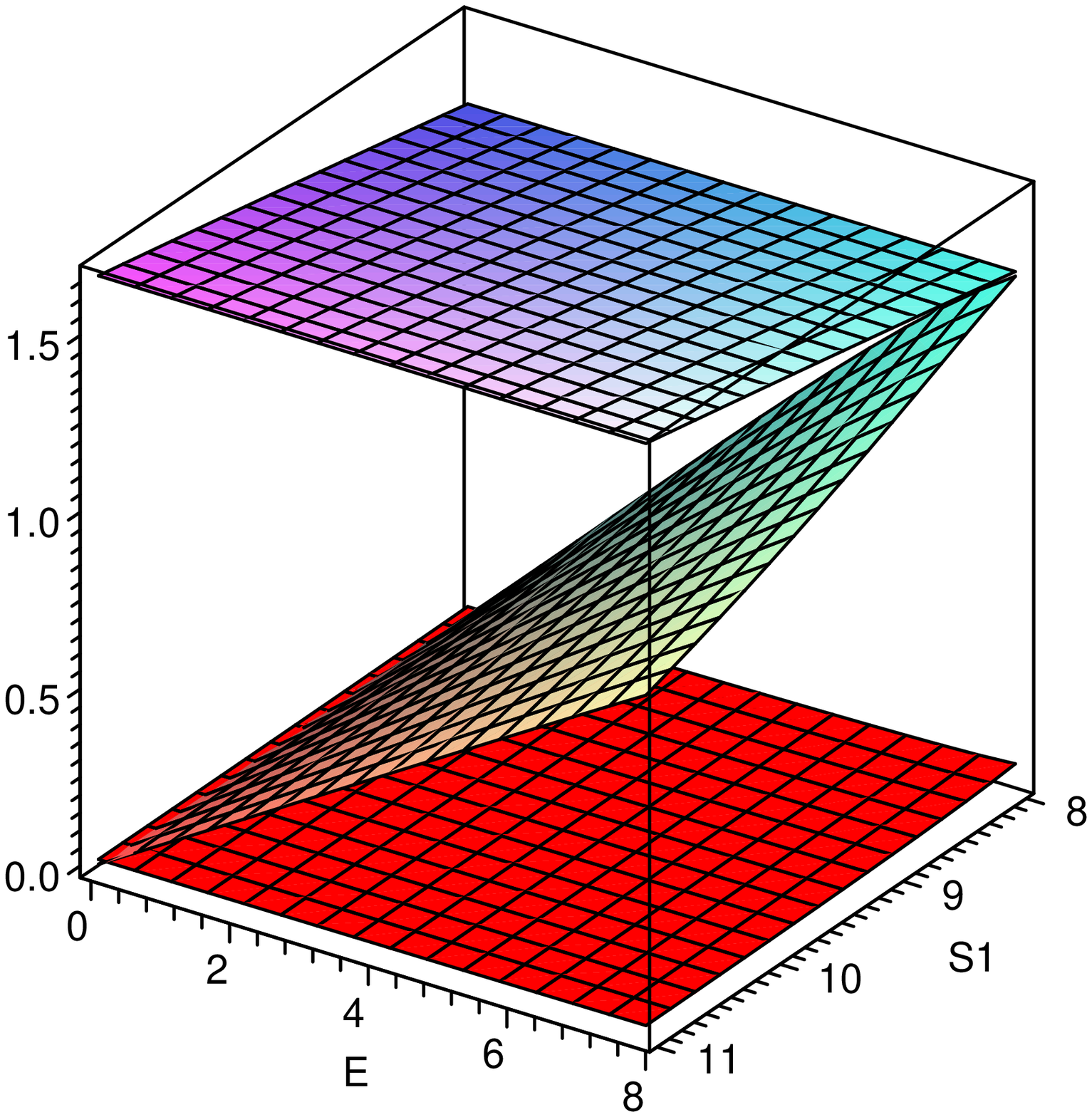}
\end{center}
\vspace{-5mm}
\caption{Similarly to 
figure \ref{fig:compare}, 
the top surface is the $\alpha'$ correction, the second is the 
$g_{\rm s}$ correction (with $F=1$ in the left graph and $F=3$ in the right), 
and the ``red carpet'' is $10/\Delta$, with $\Delta$ from \eqref{Delta2},
using the same values as in fig.\ \ref{fig:compare}. The result is qualitatively
the same as before. Note, however, that the range for 
$\Et_s$ differs. For larger values of $F$ one does not need to fine-tune $\Et_s$
as much in order for the two corrections to become of similar order. 
\label{fluxplots}}
\end{figure}

We now give the generalization of \eqref{full loop}-\eqref{Vwnpzero} 
when using the modified 
K\"ahler potential \eqref{realistic k2}.
The three contributions at leading order (${\cal O} (\V^{-3})$) 
in the large volume expansion are
\be \label{fullloop2}
V_{\mathrm{np1}}&=&e^{K_{\rm cs}} \frac{24 a^2|A|^2\tau_s^{3/2}
e^{-2a\tau_s}} {\V\Delta}\ ,\\[0.5cm]
V_{\mathrm{np2}}&=&-e^{K_{\rm cs}}  \frac{2a|AW_0|\tau_s
  e^{-a\tau_s}}{ S_1 \V^2}\left[ 1+ \frac{6\Et_s}{\Delta} \left( 1 - 2
\frac{F^2 S_1}{\tau_s} \right) \right]\ ,\\[0.5cm]
V_3&=& \frac{3 e^{K_{\rm
      cs}}|W_0|^2 } { 8 \V^3}
      \Bigg[
S_1^{1/2} \tilde{\xi} + \\
&&\hspace{-7mm} +\frac{\sqrt{\tau_s} \left(4(\Et_s)^2 
-8 (\Et_s)^2 F^2 S_1 \tau_s^{-1} (1 + F^2 S_1 \tau_s^{-1}) 
-\tfrac{8 \sqrt{2}}{3} F^2 S_1^2 \Et_s \right)}{S_1^{2}\Delta}\Bigg]\; , \nonumber 
 \label{Vwnpzero2}
\ee
where the axion has already been minimized for, as discussed in
section \ref{sec:many}, and now $\Delta$ is generalized to
\be \label{Delta2}
 \Delta&\equiv& \sqrt{2} S_1\tau_s-3 \Et_s \left(1-3\frac{ F^2 S_1}{\tau_s}\right) \ .
\ee
Plots for $F=1$ and $F=3$ are given in fig.\ \ref{fluxplots}, and 
they look quite similar to the plot without flux, fig.\ \ref{fig:compare}.
Qualitatively, the conclusion is the same; only
for nongeneric values of the $g_{\rm s}$ corrections do 
they compete with the $\alpha'$ correction. Note, however,
that the amount of fine-tuning seems to depend on the value of the 
flux, cf.\ fig.\ \ref{fluxplots}. The same is true for the 
dependence of the values of $\V$ and $\tau_s$ at the minimum
on $S_1$ and $\Et_s$. For $F=3$, for instance, 
this dependence becomes more 
complicated than what we found in \eqref{fits}. For the parameter 
range shown in fig.\ \ref{fluxplots}, 
the values of $\tau_s$ and $\V$ in the minimum vary in the ranges
$\tau_s \in [14.6, 46.3]$ and $\log_{10} \V \in [3.7, 15.5]$,
where the smallest value  for both of them is reached in the corner
where the two corrections become comparable.

Also the cancellation that we found for the gaugino masses 
survives the inclusion of the flux factor in \eqref{realistic k2}. The correction 
still only appears at sub-sub-leading order in an expansion in $\ln(1/m_{3/2})$ 
and we find (again using the dilute flux approximation for the prefactor 
$({\rm Re} f)^{-1}$):
\be  \label{subsubgeneral}
|M_s|&=&\left| \frac{F^s}{2\tau_s} \right|
=3 e^{K/2}|W_0| \left|\frac{1}{4a\tau_s}
+\frac{1}{16a^2\tau_s^2}+
\frac{S_1-12 \sqrt{2} a(1-2F^2 S_1 \tau_s^{-1}) \Et_s}{64S_1a^3\tau_s^3}
+\ldots \right| \nonumber \\[2mm]
&\sim &\frac{m_{3/2}}{{\rm ln}(1/m_{3/2})}\left[ 1+\mathcal{O}\left(
\frac{1}{{\rm ln}(1/m_{3/2}) } \right)\right]\ .
\ee
This concludes our brief study of the direct effects of fluxes
on the loop corrections.


\section{The orientifold calculation} \label{app:tori}

In the main text, we are interested in how $\Delta\! K_{g_{\rm s}}$,
the one-loop correction to the
K\"ahler potential,  scales with the
K\"ahler moduli $T_i$.
Our argument in section \ref{general} is based on the 
known result for $\Delta\! K_{g_{\rm s}}$
in the case of ${\cal N}=2$ supersymmetric $K3 \times T^2$ orientifolds 
and ${\cal N}=1$ supersymmetric $T^6/\mathbb{Z}_N$ 
(or $T^6/(\mathbb{Z}_2 \times \mathbb{Z}_2)$) orientifolds, 
from \cite{Berg:2005ja} (see also \cite{Berg:2004ek}).
Here we review this computation for the case of $K3 \times T^2$, and 
take this opportunity to adapt it to our case of D3-branes
and D7-branes from the beginning. (One can 
also obtain them by T-duality on
the final D9/D5 results of \cite{Berg:2005ja}, e.g.\ as in the
appendix of \cite{Berg:2004sj}, but as we shall
see, the direct computation is enlightening in its own right.) 
We will leave out 
details that 
are essentially identical to \cite{Berg:2005ja}, 
and only emphasize the differences. 

As shown in \cite{Berg:2005ja} using ``K\"ahler adapted'' vertex operators,
the easiest way to compute
$\Delta\! K_{g_{\rm s}}$ is by considering the 
2-point function of the complex structure modulus $U$ of $T^2$, with vanishing Wilson line 
moduli, i.e.\  
\be \label{Ucorr}
\langle V_{U} V_{\bar U} \rangle &=& -\sum_\sigma \frac{4 g_c^2 \alpha'^{-4}}{(U-\bar U)^2} 
\langle V_{ZZ}^{(0,0)}V_{\bar Z\bar Z}^{(0,0)} \rangle_\sigma\ .
\ee
Here, we use the notation of \cite{Berg:2005ja}, 
\be
V_{U}^{(0,0)} &=& - g_c \alpha'^{-2} \frac{2}{U-\bar U} V_{ZZ}^{(0,0)}\ , \non
V_{\bar U}^{(0,0)} &=& g_c \alpha'^{-2} \frac{2}{U-\bar U} V_{\bar Z \bar Z}^{(0,0)} 
\ee
and 
\be
V_{ZZ}^{(0,0)} &=& -\frac{2}{\alpha'} \int_\Sigma d^2z\
\big[i \partial Z + \frac{1}{2} 
\alpha' (p \cdot \psi) \Psi \big]
\big[i \bar \partial Z + \frac{1}{2} \alpha' (p \cdot \tilde \psi) \tilde \Psi \big]
e^{ipX} \ , \non
V_{\bar Z \bar Z}^{(0,0)} &=& -\frac{2}{\alpha'} \int_\Sigma d^2z\
\big[i \partial \bar Z + \frac{1}{2} \alpha' (p \cdot \psi) \bar \Psi \big]
\big[i \bar \partial \bar Z + \frac{1}{2} \alpha' (p \cdot \tilde \psi) 
\bar{\tilde \Psi} \big] e^{ipX}\ .
\ee
As in \cite{Berg:2005ja}, and \cite{Lust:2004cx} before that, we 
find these complex worldsheet variables particularly convenient:
\be \label{zpsi}
Z = \sqrt{\frac{\sqrt{G_{\rm str}}}{2 U_2}} (X^{4} + \bar
U X^{5}) \quad , \quad
\bar Z = \sqrt{\frac{\sqrt{G_{\rm str}}}{2 U_2}} (X^{4}
+ U X^{5})\ , \non
\Psi = \sqrt{\frac{\sqrt{G_{\rm str}}}{2 U_2}} (\psi^{4} + \bar U
\psi^{5}) \quad , \quad
\bar \Psi = \sqrt{\frac{\sqrt{G_{\rm str}}}{2 U_2}} (\psi^{4}
+ U \psi^{5})\ ,
\ee
where $\sqrt{G_{\rm str}}$ is the volume of $T^2$ measured in string frame.
The 2-point function (\ref{Ucorr}) can be expanded for small momenta,
$p_1 \cdot p_2 \ll 1$, and we obtain 
\be \label{ZZbZbZ}
\langle V_{Z Z}^{(0,0)}V_{\bar Z \bar Z}^{(0,0)} \rangle_\sigma &=&
- V_4 \frac{(p_1 \cdot p_2) \sqrt{G_{\rm str}}^{-1}}{16 (4 \pi^2 \alpha')^2}
\int_0^\infty\! 
\frac{dt}{t^4}
  \int_{\cf_\sigma} \! d^2\nu_1 d^2\nu_2 \non
&&  \hspace{-1cm} \times \sum_{k=0,1} \sum_{\vec n=(n,m)^T}  \tr
\Bigg[
 e^{-\pi \vec{n}^{T} G^{-1}_{\rm str} \vec{n} t^{-1}} 
\sum_{ \genfrac{}{}{0pt}{}{\alpha\beta} {\rm even} }
\frac{\thbw{\alpha}{\beta}(0,\tau)}{\eta^3(\tau)}
 \gamma_{\sigma,k} \cz_{\sigma,k}^{\rm int} \zba{\alpha}{\beta} \non
&& \hspace{-1cm}
 \times
\Big[ \langle \bar \partial Z(\bar \nu_1)
\bar \partial \bar{Z}(\bar \nu_2) \rangle_\sigma \langle \Psi(\nu_1)
\bar \Psi(\nu_2) \rangle_\sigma^{\alpha, \beta} \langle \psi(\nu_1)
\psi(\nu_2) \rangle_\sigma^{\alpha, \beta} \\
&& \hspace{-8mm}
+ \langle \bar \partial Z(\bar \nu_1)
\partial \bar Z(\nu_2) \rangle_\sigma \langle \Psi(\nu_1)
\bar{\tilde{\Psi}}(\bar \nu_2) \rangle_\sigma^{\alpha, \beta} \langle \psi(\nu_1) \tilde
\psi(\bar \nu_2) \rangle_\sigma^{\alpha, \beta} + {\rm c.c.} \Big]
\Bigg] + {\cal O} ((p_1 \cdot p_2)^2)\ . \nonumber
\ee
For the details we refer to \cite{Berg:2005ja}. The main difference to the 
corresponding formula (C.3) in
\cite{Berg:2005ja} is the appearance of the inverse metric $G_{\rm str}^{-1}$ 
in the exponent arising from the zero mode sum, and in the prefactor. This is due to the 
fact that the D3 and D7 branes are localized along the $T^2$, and
so the closed string channel involves a Kaluza-Klein 
momentum sum instead of a winding sum. 
The sum over bosonic zero modes has been made explicit, 
since there is also an implicit dependence on $m,n$ 
in the bosonic correlators: this arises from the
classical piece in the split into zero modes and fluctuations. 
That is, $Z(\nu) = Z_{\rm class}(\nu) +
Z_{\rm qu}(\nu)$, where the classical part is given by
\be \label{zeromode}
Z_{\rm class} = \sqrt{\alpha'} \sqrt{\frac{\sqrt{G_{\rm str}}}{2 U_2}} \Big(n + m
\bar U \Big) {\rm Re}(\nu) \, \tilde c_\sigma\ , \quad \tilde c_\sigma = \left\{
\begin{array}{ll}
1& {\rm for}\ \ck  \\
2&{\rm for}\  \ca \ , \cm
\end{array} \right.\ .
\ee
These zero modes have the right periodicity under ${\rm Re}(\nu)
\rightarrow {\rm Re}(\nu) + \pi$ (for $\ca , \cm$) or ${\rm
Re}(\nu) \rightarrow {\rm Re}(\nu) + 2 \pi$ (for $\ck$), 
i.e.\ $X^4 \rightarrow X^4 + 2 \pi n \sqrt{\alpha'}$ and $X^5
\rightarrow X^5 + 2 \pi m \sqrt{\alpha'}$. In contrast to \cite{Berg:2005ja}
they involve the real part of $\nu$. The reason is again that in the D3/D7 case the branes are 
localized along $T^2$ and thus the winding appears in the open string channel 
as opposed to the closed string channel (as was the case for D9/D5 branes). 

The sum over spin structures is performed using Riemann identities.
This leaves the correlators of the bosonic fields as the only
piece that depends on the positions $\nu_i$ of the vertex
operators. The $\nu_i$ integral can then be evaluated.
As the zero modes (\ref{zeromode}) involve the real part of 
$\nu$ in the case of D3/D7-branes, in
contrast to the D9/D5-case studied in \cite{Berg:2005ja}, the zero mode contribution in the 
$Z$-correlators drops out. The 
quantum part is evaluated using the method of images
on the worldsheet \cite{Burgess:1986ah,Burgess:1986wt,Antoniadis:1996vw}. 
To evaluate the KK sum in \eqref{ZZbZbZ}, it is useful to regularize the
integral over $t$ by a UV cutoff $\Lambda$. With this we obtain
\be \label{intKK}
&& \int_{1/(e_\sigma \Lambda^2)}^\infty  \frac{dt}{t^4}
\sum_{\vec n=(n,m)^T}\!\!\!\!\!\! ' \; \; \;  \Bigg[ 
\pi^3 c_\sigma^2 t \alpha'  e^{-\pi \vec{n}^{T} G^{-1}_{\rm str} \vec{n} t^{-1}} \Bigg] \non
&&
\hspace{0cm} = \frac12 \pi^3 \alpha'c_\sigma^2 e_\sigma^2 \Lambda^4  +
\pi \alpha' c_\sigma^2\sqrt{G_{\rm str}}^2 E_2 (0, U) ~+~ \ldots \ ,
\ee
where the prime at the sum indicates that the  $(n,m)=(0,0)$ term 
is left out, and $c_{\sigma}$, $e_{\sigma}$ are constants
whose precise values will not be important in the following (but 
can be found in \cite{Berg:2005ja}).
Terms that go to zero in the limit $\Lambda\rightarrow \infty$ have
been dropped, as indicated by the ellipsis.
The nonholomorphic Eisenstein series $E_2(0,U)$ is 
the $s=2$ special case of
\be \label{eisen}
E_s (0, U) &=& \sum_{\vec n=(n,m)^T} \!\!\!\!\!\! ' \; \; \; \frac{U_2^s}{|n+mU|^{2s}}\ .
 \ee
The terms involving the UV cutoff $\Lambda$ drop out after summing
over all diagrams, due to tadpole cancellation. We
have then reduced \eqref{ZZbZbZ} to
\be \label{final}
\langle V_{Z Z}^{(0,0)}V_{\bar Z\bar Z}^{(0,0)} \rangle_\sigma
&=&-(p_1 \cdot p_2) \alpha' \frac{V_4}{(4 \pi^2 \alpha')^2} 
\frac{c_\sigma^2 \pi \sqrt{G_{\rm str}}}{8}
\sum_{k} \tr \Big[ E_2(0,U) \gamma_{\sigma,k}
\cq_{\sigma,k} \Big] \nonumber \\
&&\hspace{6cm}  + {\cal O} ((p_1 \cdot p_2)^2) \ . 
\ee
The quantities $\cq_{\sigma,k}$ come from the sum over spin structures and
are defined in \cite{Berg:2005ja}. Introducing the notation 
\be \label{e2au_def}
\ce_2(0,U) = \sum_{\sigma} c_\sigma^2
\sum_{k=0,1} \tr \Big[
E_2(0,U) \gamma_{\sigma,k} \cq_{\sigma,k} \Big] \ ,
\ee
we end up with (neglecting some irrelevant factors 
of $g_c, \alpha'$, terms subleading in the low-energy expansion,
and constants of order 1)
\be \label{s2stringframe}
\langle V_{U} V_{\bar U} \rangle \sim - i (p_1 \cdot p_2) 
\frac{V_4}{(4 \pi^2 \alpha')^2} 
\frac{\sqrt{G_{\rm str}}}{(U-\bar U)^2}  \ce_2(0,U)\ .
\ee
To read off the one-loop correction to the kinetic
term of $U$ we need to perform a Weyl rescaling to the Einstein
frame. In the one-loop term (\ref{s2stringframe}) this just leads
to
\be \label{Weyl}
\mbox{Weyl rescaling:} \quad \times \frac{e^{2 \Phi}}{\cv^{\rm str}} \ ,
\ee
where 
\be \label{volumes}
\cv^{\rm str} =  \cv^{\rm str}_{\rm K3} \sqrt{G_{\rm str}}
\ee
is the overall volume in string frame.
The K\"ahler potential can then be read off from the kinetic term by
use of
the identity
\be \label{derivE}
\partial_U \partial_{\bar U} E_2(0,U) = - \frac{2}{(U-\bar U)^2} E_2(0,U)\ ,
\ee
producing the final result
\be \label{k1} 
\Delta\! K_{g_{\rm s}} \; \sim \; 
\frac{\sqrt{G_{\rm str}}e^{\Phi}}{\cv^{\rm str} (S+\bar S)}  \ce_2(0,U)\ ,
\ee
where $\sqrt{G_{\rm str}}e^{\Phi}/\cv^{\rm str}$
is to be interpreted as a function of the K\"ahler variables. 
In the $K3 \times T^2$ orientifold case, using (\ref{volumes}), 
this is just proportional to
$e^{\Phi}/ \cv^{\rm str}_{\rm K3} \sim (T+\bar T)^{-1}$ (with ${\rm Re}\ T$
the volume of $K3$ measured in Einstein frame), 
giving a result T-dual to \cite{Berg:2005ja} (note that we switched 
the real and imaginary parts in the definition of $T$ and $S$ 
as compared to \cite{Berg:2005ja}, to conform with the rest of this paper).
As we argue in section \ref{general},
in general the dependence on the
K\"ahler moduli will be more complicated than this, because there 
is no analog to the relation (\ref{volumes}). 
It is still clear that the inverse suppression in the overall volume 
will appear as in (\ref{k1}), given that it is a direct consequence of the
Weyl rescaling.


\section{Factorized approximation}
\label{calcu}

As mentioned in section \ref{sec:othersoft}, 
it is an important issue to what extent the moduli spaces
of K\"ahler and complex structure moduli factorize.
In this appendix, we give further details on the factorized
approximation.

A common starting point in the analysis of the potential arising in type IIB theory 
with 3-form fluxes is to assume
 that all complex structure moduli $U^\alpha$ and the dilaton $S$
are stabilized by demanding 
\be \label{dusw}
D_{U^\alpha} W = 0 = D_S W\ .  
\ee
In this case the F-term potential for the moduli (\ref{Fterm}) reduces to
\be \label{onlykaehler}
V=e^{K}\left( G^{\bar{\jmath}i}D_{\bar{\jmath}}\bar{W}D_iW -3|W|^2\right)\ ,
\ee
where as in the main text,
the indices $i$ and $j$ refer only to the K\"ahler moduli and thus run from $1$ to 
$h^{1,1}$. Note that 
even though the complex structure moduli and the dilaton are assumed 
to be stabilized by (\ref{dusw}), the inverse metric $G^{\bar{\jmath}i}$ 
is part of the inverse 
of the whole moduli space metric. More precisely, if we
denote the K\"ahler moduli by $T_i$, as before, and all other moduli (i.e.\
the complex structure moduli and the dilaton) collectively as $Z^a$,
the  moduli space 
metric is given by
\be\label{block}
G_{I \bar J} &\sim& \left( \begin{array}{cc} K_{i \bar{\jmath}} &         K_{i \bar b} \\
                                           K_{a \bar{\jmath}}&         K_{a\bar{b}}
                                   \end{array} 
                          \right)\ . 
\ee
We denote the inverse of this (whole) metric by $G^{\bar J I}$. In general
\be \label{kij}
G^{\bar \jmath i} \neq (K_{i \bar{\jmath}})^{-1}\ .
\ee
Equality only holds if $G_{i \bar b}=0$, i.e.\ if the moduli space of the K\"ahler moduli
is factorized from the rest, as it is the case without loop and
$\alpha'$ corrections.

In this appendix, we would like to investigate at which order in a large volume 
expansion the two matrices in (\ref{kij}) start to deviate from each other. 
For this analysis we assume a volume of the ``Swiss cheese'' form as in \eqref{volume}
and a K\"ahler potential of the form (\ref{LVSLoop}) (without
taking possible effects of fluxes on the KK and winding mode 
spectra into account as was done in appendix 
\ref{E7neq0}; thus, $\gk^{\mathfrak{a}} \sim t_{\mathfrak{a}}$
and $\gk^{\mathfrak{q}} \sim t^{-1}_{\mathfrak{q}}$ for some 
2-cycle volumes). To avoid 
cumbersome notation we will indicate all the small moduli collectively 
as $\tau_s$. We then use the formula
\be \label{schur}
 \left( \begin{array}{cc} A & B \\
                        C & D
                                   \end{array} 
                          \right)^{-1} &=& 
                           \left( \begin{array}{cc} A^{-1}(1+BP^{-1}CA^{-1}) & -A^{-1}BP^{-1} \\
                                                   -P^{-1}CA^{-1} & P^{-1} 
                                   \end{array} 
                          \right) \ ,
\ee
where $P$ is the Schur complement of $A$, defined as
\be
P=D-CA^{-1}B\ .
\ee
In our case $P$ is the Schur complement of $K_{i\bar{\jmath}}$. 
From \eqref{LVSLoop} we read off that
\be \label{scaling}
G_{I\bar J}\sim\left( 
        \begin{array}{cccc}
                \tau_b^{-2} & \tau_b^{\expo} & \tau_b^{-2}    & \tau_b^{-2} \\
              \tau_b^{ \expo}    & \tau_b^{-3/2} & \tau_b^{-3/2}  & \tau_b^{-3/2} \\
              \tau_b^{     -2}& \tau_b^{-3/2}     & \tau_b^{0}     & \tau_b^{-1} \\
               \tau_b^{    -2}& \tau_b^{     -3/2}&\tau_b^{-1}       & \tau_b^{0}
        \end{array}
        \right)\ ,
\ee
where we only indicate the $\tau_b$ dependence and the indices run over 
$I,J=\{T_b,T_s,U,S\}$. Here $\expo=-2$ for those $\tau_{i}$ with a 
nonvanishing $a_i$ in \eqref{volume} (so $\expo$ has an implicit index $i$),
otherwise $\expo=-5/2$ (which is in particular the value in the 
$\mathbb{P}_{[1,1,1,6,9]}$ case). We decompose $G_{I \bar J}$ as in \eqref{schur}
\be 
A\sim \left( 
        \begin{array}{cc}
                \tau_b^{-2 }  &\tau_b^{ \expo }\\
               \tau_b^{ \expo} &\tau_b^{ -3/2}     
        \end{array} \right)\ , 
& &   A^{-1}\sim 
        \left( 
        \begin{array}{cc}
                \tau_b^{2 }&\tau_b^{ 7/2+\expo} \\
                \tau_b^{7/2+\expo }& \tau_b^{3/2}         
        \end{array} \right)\ , \nonumber \\
        B=C^{T}\sim
        \left( 
        \begin{array}{cc}
                \tau_b^{-2  } & \tau_b^{-2} \\
               \tau_b^{ -3/2} & \tau_b^{-3/2}     
        \end{array} \right)\ , 
&&
        D\sim \left( 
        \begin{array}{cc}
                \tau_b^{0  }&\tau_b^{ -1 }\\
                \tau_b^{-1} &\tau_b^{ 0  }
        \end{array} \right)\ , \nonumber \\
        P\sim \left( 
        \begin{array}{cc}
                \tau_b^{0  }&\tau_b^{ -1 }\\
                \tau_b^{-1 }&\tau_b^{ 0  }
        \end{array} \right)\ ,
&&
        P^{-1}\sim \left( 
        \begin{array}{cc}
                \tau_b^{0  }&\tau_b^{ -1 }\\
                \tau_b^{-1 }&\tau_b^{ 0  }
        \end{array} \right)\ .
\ee
Using \eqref{schur} one easily obtains the scaling of the inverse:
\be 
G^{\bar J I}\sim\left( 
        \begin{array}{cccc}
                 \tau_b^{2 }&\tau_b^{ 7/2+\expo}   & \tau_b^{0}  & \tau_b^{0} \\
            \tau_b^{7/2+\expo}       & \tau_b^{3/2} & \tau_b^{0}  & \tau_b^{0} \\
                   \tau_b^{0}&\tau_b^{0}     &\tau_b^{ 0}  &\tau_b^{ -1 }\\
                   \tau_b^{0} &\tau_b^{0}     &\tau_b^{-1}    &\tau_b^{ 0}
        \end{array}
        \right)\ .
\ee
Now, from \eqref{schur}, $G^{\bar \jmath i}$ receives two contributions. The first 
is $K^{\bar \jmath i}$, that would be the only term in the case of a factorized 
metric; the second is $K_{h\bar{\jmath}}^{-1}\,K_{h\bar b}P^{-1}_{a\bar b}
K_{a\bar{l} }K_{i\bar{l}}^{-1}$, that  
breaks factorization. Let us compare their $\tau_b$ scaling:
\be \label{sup}
G^{\bar \jmath i}&=& A^{-1}+A^{-1}BP^{-1}CA^{-1} \\
&\sim&\left( 
        \begin{array}{cc}
               \tau_b^{ 2} &\tau_b^{ 7/2+\expo} \\
               \tau_b^{ 7/2+\expo} & \tau_b^{3/2}         
        \end{array} \right) \nonumber
        + \left( 
        \begin{array}{cc}
                \tau_b^{0} &\tau_b^{ 0 }\\
                \tau_b^{0 }&\tau_b^{ 0}   
        \end{array} \right)\ .
\ee
Thus the corrections coming from non-vanishing off-diagonal metric elements in 
(\ref{block}) set in with a suppression by $\tau_b^{-2}$, $\tau_b^{-7/2-\expo}$ 
and $\tau_b^{-3/2}$ in $G^{\bar b b}$, $G^{\bar b s}$ and $G^{\bar s s}$, respectively. 
In the explicit example based on $\mathbb{P}_{[1,1,1,6,9]}$, $\expo=-5/2$, and we checked this 
result by comparing to the subleading terms in \eqref{inv metric}.


\subsection{Factorized approximation of the scalar potential}

What we are really interested in is not the (inverse) metric itself, but the 
scalar potential, to which we now turn. For the nonperturbative 
terms $V_{\rm np1}$ and $V_{\rm np2}$, the suppression of the off-diagonal terms in 
\eqref{sup} is inherited by the scalar potential, as they are proportional 
to $G^{\bar s s} \bar{W}_{,\bar s} W_{,s}$ and $G^{\bar \jmath s} K_{\bar \jmath}$,
respectively. For $V_3$ things are not as simple, due to the
no-scale structure at leading order. Let us neglect for a 
moment all the quantum corrections, then the no-scale structure implies
\be 
\left[G^{\bar{\imath} j}K_{\bar{\imath}}K_{j}-3\right]_{\rm no-scale}&\sim&
(\tau_b^{-1},\tau_b^{\expo+1})
\left( 
        \begin{array}{cc}
               \tau_b^{ 2} &\tau_b^{ 7/2+\expo} \\
               \tau_b^{ 7/2+\expo} & \tau_b^{3/2}         
        \end{array} \right) \nonumber
	\left( 
        \begin{array}{c}
               \tau_b^{ -1} \\
	       \tau_b^{\expo+1}
        \end{array} \right)-3 \nonumber \\
	&\sim&\tau_b^{0}+\tau_b^{2\expo+7/2}=0\ .
\ee
The two terms have to vanish independently. Now let us add corrections
that break
no-scale structure. Because of the cancellation described 
in appendix \ref{canc ap}, the leading contribution can be seen to 
come at order $\tau_b^{-3/2}$ (from the $\al$, $\Et_s$ and $\Es_s$ 
corrections). On the other hand, the off-diagonal terms appear at order
\be \label{off diag pot}
\left[G^{\bar{\imath} j}K_{\bar{\imath}}K_{j}\right]_{\rm off-diagonal}
&\sim&(\tau_b^{-1},\tau_b^{\expo+1})
\left( 
        \begin{array}{cc}
               \tau_b^{ 0} &\tau_b^{0} \\
               \tau_b^{ 0} & \tau_b^{0}         
        \end{array} \right) \nonumber
	\left( 
        \begin{array}{c}
               \tau_b^{ -1} \\
	       \tau_b^{\expo+1}
        \end{array} \right) \nonumber \\
	&\sim&\tau_b^{-2}+\tau_b^{\expo}+\tau_b^{2\expo+2}\nonumber\\
	&\sim&\tau_b^{-2}+\dots\ ,
\ee
for both $\expo=-2$ and $\beta=-5/2$. 
Therefore, the off-diagonal terms of the moduli space 
metric appear in the scalar potential with a suppression of at least $\tau_b^{-1/2}$ 
(as is confirmed by the explicit example of section \ref{P model}, cf.\
formulas \eqref{full loop}-\eqref{e31}). The suppression can be 
even stronger if some corrections are absent and 
the leading term in \eqref{off diag pot} vanishes.

To summarize: if one is only interested in the leading term  of the scalar potential 
in the large volume (i.e.\ large $\tau_b$) expansion, 
then one can use the \textit{factorized 
approximation}, i.e.\ 
\be 
 G^{\bar \jmath i}=K^{\bar \jmath i}+ \mathcal{O}\left( \tau_b^0 \right).
\ee
This provides a useful tool to simplify the calculations.


\bibliographystyle{Jopt2}
\bibliography{gg3}

\providecommand{\href}[2]{#2}\begingroup\raggedright\begin{thebibliography}{10}

\bibitem{Giddings:2001yu}
S.~B. Giddings, S.~Kachru, and J.~Polchinski, { Hierarchies from fluxes in
  string compactifications},  { Phys. Rev.} { D66} (2002) 106006,
[\href{http://xxx.lanl.gov/abs/hep-th/0105097}{{\tt hep-th/0105097}}].

\bibitem{Kachru:2003aw}
S.~Kachru, R.~Kallosh, A.~Linde, and S.~P. Trivedi, { De {S}itter vacua in
  string theory},  { Phys. Rev.} { D68} (2003) 046005,
[\href{http://xxx.lanl.gov/abs/hep-th/0301240}{{\tt hep-th/0301240}}].

\bibitem{Balasubramanian:2005zx}
V.~Balasubramanian, P.~Berglund, J.~P. Conlon, and F.~Quevedo, { Systematics of
  moduli stabilisation in {C}alabi-{Y}au flux compactifications},  { JHEP} {
  03} (2005) 007,
[\href{http://xxx.lanl.gov/abs/hep-th/0502058}{{\tt hep-th/0502058}}].

\bibitem{Conlon:2005ki}
J.~P. Conlon, F.~Quevedo, and K.~Suruliz, { Large-volume flux
  compactifications: Moduli spectrum and {D3}/{D7} soft supersymmetry
  breaking},  { JHEP} { 08} (2005) 007,
[\href{http://xxx.lanl.gov/abs/hep-th/0505076}{{\tt hep-th/0505076}}].

\bibitem{Becker:2002nn}
K.~Becker, M.~Becker, M.~Haack, and J.~Louis, { Supersymmetry breaking and
  alpha'-corrections to flux induced potentials},  { JHEP} { 06} (2002) 060,
[\href{http://xxx.lanl.gov/abs/hep-th/0204254}{{\tt hep-th/0204254}}].

\bibitem{Conlon:2006us}
J.~P. Conlon and F.~Quevedo, { Gaugino and scalar masses in the landscape},  {
  JHEP} { 06} (2006) 029,
[\href{http://xxx.lanl.gov/abs/hep-th/0605141}{{\tt hep-th/0605141}}].

\bibitem{Conlon:2006wz}
J.~P. Conlon, S.~S. Abdussalam, F.~Quevedo, and K.~Suruliz, { Soft {SUSY}
  breaking terms for chiral matter in {IIB} string compactifications},  { JHEP}
  { 01} (2007) 032,
[\href{http://xxx.lanl.gov/abs/hep-th/0610129}{{\tt hep-th/0610129}}].

\bibitem{Conlon:2006tq}
J.~P. Conlon, { The {QCD} axion and moduli stabilisation},  { JHEP} { 05}
  (2006) 078,
[\href{http://xxx.lanl.gov/abs/hep-th/0602233}{{\tt hep-th/0602233}}].

\bibitem{Conlon:2006ur}
J.~P. Conlon, { Seeing the invisible axion in the sparticle spectrum},  { Phys.
  Rev. Lett.} { 97} (2006) 261802,
[\href{http://xxx.lanl.gov/abs/hep-ph/0607138}{{\tt hep-ph/0607138}}].

\bibitem{Conlon:2006wt}
J.~P. Conlon and D.~Cremades, { The neutrino suppression scale from large
  volumes},
\href{http://xxx.lanl.gov/abs/hep-ph/0611144}{{\tt hep-ph/0611144}}.

\bibitem{Conlon:2005jm}
J.~P. Conlon and F.~Quevedo, { Kaehler moduli inflation},  { JHEP} { 01} (2006)
  146,
[\href{http://xxx.lanl.gov/abs/hep-th/0509012}{{\tt hep-th/0509012}}].

\bibitem{Holman:2006ek}
R.~Holman and J.~A. Hutasoit, { Axionic inflation from large volume flux
  compactifications},
\href{http://xxx.lanl.gov/abs/hep-th/0603246}{{\tt hep-th/0603246}}.

\bibitem{Simon:2006du}
J.~Simon, R.~Jimenez, L.~Verde, P.~Berglund, and V.~Balasubramanian, { Using
  cosmology to constrain the topology of hidden dimensions},
\href{http://xxx.lanl.gov/abs/astro-ph/0605371}{{\tt astro-ph/0605371}}.

\bibitem{Bond:2006nc}
J.~R. Bond, L.~Kofman, S.~Prokushkin, and P.~M. Vaudrevange, { Roulette
  inflation with {K}aehler moduli and their axions},
\href{http://xxx.lanl.gov/abs/hep-th/0612197}{{\tt hep-th/0612197}}.

\bibitem{Kane:2006yi}
G.~L. Kane, P.~Kumar, and J.~Shao, { {LHC} string phenomenology},
\href{http://xxx.lanl.gov/abs/hep-ph/0610038}{{\tt hep-ph/0610038}}.

\bibitem{Denef:2004dm}
F.~Denef, M.~R. Douglas, and B.~Florea, { Building a better racetrack},  {
  JHEP} { 06} (2004) 034,
[\href{http://xxx.lanl.gov/abs/hep-th/0404257}{{\tt hep-th/0404257}}].

\bibitem{Gorlich:2004qm}
L.~G{\"o}rlich, S.~Kachru, P.~K. Tripathy, and S.~P. Trivedi, { Gaugino
  condensation and nonperturbative superpotentials in flux compactifications},
  { JHEP} { 12} (2004) 074,
[\href{http://xxx.lanl.gov/abs/hep-th/0407130}{{\tt hep-th/0407130}}].

\bibitem{Tripathy:2005hv}
P.~K. Tripathy and S.~P. Trivedi, { {D3} brane action and fermion zero modes in
  presence of background flux},  { JHEP} { 06} (2005) 066,
[\href{http://xxx.lanl.gov/abs/hep-th/0503072}{{\tt hep-th/0503072}}].

\bibitem{Denef:2005mm}
F.~Denef, M.~R. Douglas, B.~Florea, A.~Grassi, and S.~Kachru, { Fixing all
  moduli in a simple {F}-theory compactification},  { Adv. Theor. Math. Phys.}
  { 9} (2005) 861--929,
[\href{http://xxx.lanl.gov/abs/hep-th/0503124}{{\tt hep-th/0503124}}].

\bibitem{Saulina:2005ve}
N.~Saulina, { Topological constraints on stabilized flux vacua},  { Nucl.
  Phys.} { B720} (2005) 203--210,
[\href{http://xxx.lanl.gov/abs/hep-th/0503125}{{\tt hep-th/0503125}}].

\bibitem{Kallosh:2005gs}
R.~Kallosh, A.-K. Kashani-Poor, and A.~Tomasiello, { Counting fermionic zero
  modes on {M5} with fluxes},  { JHEP} { 06} (2005) 069,
[\href{http://xxx.lanl.gov/abs/hep-th/0503138}{{\tt hep-th/0503138}}].

\bibitem{Martucci:2005rb}
L.~Martucci, J.~Rosseel, D.~Van~den Bleeken, and A.~Van~Proeyen, { Dirac
  actions for {D}-branes on backgrounds with fluxes},  { Class. Quant. Grav.} {
  22} (2005) 2745--2764,
[\href{http://xxx.lanl.gov/abs/hep-th/0504041}{{\tt hep-th/0504041}}].

\bibitem{Berglund:2005dm}
P.~Berglund and P.~Mayr, { Non-perturbative superpotentials in {F}-theory and
  string duality},
\href{http://xxx.lanl.gov/abs/hep-th/0504058}{{\tt hep-th/0504058}}.

\bibitem{Bergshoeff:2005yp}
E.~Bergshoeff, R.~Kallosh, A.-K. Kashani-Poor, D.~Sorokin, and A.~Tomasiello, {
  An index for the {D}irac operator on {D}3 branes with background fluxes},  {
  JHEP} { 10} (2005) 102,
[\href{http://xxx.lanl.gov/abs/hep-th/0507069}{{\tt hep-th/0507069}}].

\bibitem{Lust:2005cu}
D.~L{\"u}st, S.~Reffert, W.~Schulgin, and P.~K. Tripathy, { Fermion zero modes
  in the presence of fluxes and a non- perturbative superpotential},  { JHEP} {
  08} (2006) 071,
[\href{http://xxx.lanl.gov/abs/hep-th/0509082}{{\tt hep-th/0509082}}].

\bibitem{Lust:2006zg}
D.~L{\"u}st, S.~Reffert, E.~Scheidegger, W.~Schulgin, and S.~Stieberger, {
  Moduli stabilization in type {IIB} orientifolds. {II}},  { Nucl. Phys.} {
  B766} (2007) 178--231,
[\href{http://xxx.lanl.gov/abs/hep-th/0609013}{{\tt hep-th/0609013}}].

\bibitem{Blumenhagen:2006xt}
R.~Blumenhagen, M.~Cvetic, and T.~Weigand, { Spacetime instanton corrections in
  {4D} string vacua - the seesaw mechanism for {D}-brane models},  { Nucl.
  Phys.} { B771} (2007) 113--142,
[\href{http://xxx.lanl.gov/abs/hep-th/0609191}{{\tt hep-th/0609191}}].

\bibitem{Haack:2006cy}
M.~Haack, D.~Krefl, D.~L{\"u}st, A.~Van~Proeyen, and M.~Zagermann, { Gaugino
  condensates and {D}-terms from {D7}-branes},  { JHEP} { 01} (2007) 078,
[\href{http://xxx.lanl.gov/abs/hep-th/0609211}{{\tt hep-th/0609211}}].

\bibitem{Akerblom:2006hx}
N.~Akerblom, R.~Blumenhagen, D.~L{\"u}st, E.~Plauschinn, and
  M.~Schmidt-Sommerfeld, { Non-perturbative {SQCD} superpotentials from string
  instantons},  { JHEP} { 04} (2007) 076,
[\href{http://xxx.lanl.gov/abs/hep-th/0612132}{{\tt hep-th/0612132}}].

\bibitem{Tsimpis:2007sx}
D.~Tsimpis, { Fivebrane instantons and {Calabi-Yau} fourfolds with flux},  {
  JHEP} { 03} (2007) 099,
[\href{http://xxx.lanl.gov/abs/hep-th/0701287}{{\tt hep-th/0701287}}].

\bibitem{Bianchi:2007fx}
M.~Bianchi and E.~Kiritsis, { Non-perturbative and flux superpotentials for
  {T}ype {I} strings on the {Z}${}_3$ orbifold},
\href{http://xxx.lanl.gov/abs/hep-th/0702015}{{\tt hep-th/0702015}}.

\bibitem{Argurio:2007vq}
R.~Argurio, M.~Bertolini, G.~Ferretti, A.~Lerda, and C.~Petersson, { Stringy
  instantons at orbifold singularities},
\href{http://xxx.lanl.gov/abs/arXiv:0704.0262 [hep-th]}{{\tt arXiv:0704.0262
  [hep-th]}}.

\bibitem{Choi:2004sx}
K.~Choi, A.~Falkowski, H.~P. Nilles, M.~Olechowski, and S.~Pokorski, {
  Stability of flux compactifications and the pattern of supersymmetry
  breaking},  { JHEP} { 11} (2004) 076,
[\href{http://xxx.lanl.gov/abs/hep-th/0411066}{{\tt hep-th/0411066}}].

\bibitem{Lust:2005dy}
D.~L{\"u}st, S.~Reffert, W.~Schulgin, and S.~Stieberger, { Moduli stabilization
  in type {IIB} orientifolds. {I}: Orbifold limits},  { Nucl. Phys.} { B766}
  (2007) 68--149,
[\href{http://xxx.lanl.gov/abs/hep-th/0506090}{{\tt hep-th/0506090}}].

\bibitem{Krefl:2006vu}
D.~Krefl and D.~L{\"u}st, { On supersymmetric minkowski vacua in {IIB}
  orientifolds},  { JHEP} { 06} (2006) 023,
[\href{http://xxx.lanl.gov/abs/hep-th/0603166}{{\tt hep-th/0603166}}].

\bibitem{Gomez-Reino:2006dk}
M.~Gomez-Reino and C.~A. Scrucca, { Locally stable non-supersymmetric
  {M}inkowski vacua in supergravity},  { JHEP} { 05} (2006) 015,
[\href{http://xxx.lanl.gov/abs/hep-th/0602246}{{\tt hep-th/0602246}}].

\bibitem{Berg:2005ja}
M.~Berg, M.~Haack, and B.~K{\"o}rs, { String loop corrections to {K}aehler
  potentials in orientifolds},  { JHEP} { 11} (2005) 030,
[\href{http://xxx.lanl.gov/abs/hep-th/0508043}{{\tt hep-th/0508043}}].

\bibitem{Antoniadis:1996vw}
I.~Antoniadis, C.~Bachas, C.~Fabre, H.~Partouche, and T.~R. Taylor, { Aspects
  of type {I} - type {II} - heterotic triality in four dimensions},  { Nucl.
  Phys.} { B489} (1997) 160--178,
[\href{http://xxx.lanl.gov/abs/hep-th/9608012}{{\tt hep-th/9608012}}].

\bibitem{Angelantonj:2002ct}
C.~Angelantonj and A.~Sagnotti, { Open strings},  { Phys. Rept.} { 371} (2002)
  1--150,
[\href{http://xxx.lanl.gov/abs/hep-th/0204089}{{\tt hep-th/0204089}}].

\bibitem{Burgess:2006mn}
C.~P. Burgess, P.~Camara, S.~de~Alwis, S.~Giddings, A.~Maharana, F.~Quevedo,
  and K.~Suruliz, { Warped supersymmetry breaking},
\href{http://xxx.lanl.gov/abs/hep-th/0610255}{{\tt hep-th/0610255}}.

\bibitem{Giddings:2005ff}
S.~B. Giddings and A.~Maharana, { Dynamics of warped compactifications and the
  shape of the warped landscape},  { Phys. Rev.} { D73} (2006) 126003,
[\href{http://xxx.lanl.gov/abs/hep-th/0507158}{{\tt hep-th/0507158}}].

\bibitem{Burgess:2003ic}
C.~P. Burgess, R.~Kallosh, and F.~Quevedo, { de {S}itter string vacua from
  supersymmetric {D}-terms},  { JHEP} { 10} (2003) 056,
[\href{http://xxx.lanl.gov/abs/hep-th/0309187}{{\tt hep-th/0309187}}].

\bibitem{Lebedev:2006qq}
O.~Lebedev, H.~P. Nilles, and M.~Ratz, { de {S}itter vacua from matter
  superpotentials},  { Phys. Lett.} { B636} (2006) 126,
[\href{http://xxx.lanl.gov/abs/hep-th/0603047}{{\tt hep-th/0603047}}].

\bibitem{Dudas:2006vc}
E.~Dudas and Y.~Mambrini, { Moduli stabilization with positive vacuum energy},
  { JHEP} { 10} (2006) 044,
[\href{http://xxx.lanl.gov/abs/hep-th/0607077}{{\tt hep-th/0607077}}].

\bibitem{Dudas:2006gr}
E.~Dudas, C.~Papineau, and S.~Pokorski, { Moduli stabilization and uplifting
  with dynamically generated {F}-terms},  { JHEP} { 02} (2007) 028,
[\href{http://xxx.lanl.gov/abs/hep-th/0610297}{{\tt hep-th/0610297}}].

\bibitem{Cremmer:1983bf}
E.~Cremmer, S.~Ferrara, C.~Kounnas, and D.~V. Nanopoulos, { Naturally vanishing
  cosmological constant in {N=1} supergravity},  { Phys. Lett.} { B133} (1983)
61.

\bibitem{Candelas:1994hw}
P.~Candelas, A.~Font, S.~H. Katz, and D.~R. Morrison, { Mirror symmetry for two
  parameter models. 2},  { Nucl. Phys.} { B429} (1994) 626--674,
[\href{http://xxx.lanl.gov/abs/hep-th/9403187}{{\tt hep-th/9403187}}].

\bibitem{Curio:2006ea}
G.~Curio and V.~Spillner, { On the modified {KKLT} procedure: A case study for
  the {P}(11169)(18) model},
\href{http://xxx.lanl.gov/abs/hep-th/0606047}{{\tt hep-th/0606047}}.

\bibitem{Dine:1985he}
M.~Dine and N.~Seiberg, { Is the superstring weakly coupled?},  { Phys. Lett.}
  { B162} (1985)
299.

\bibitem{Balasubramanian:2004uy}
V.~Balasubramanian and P.~Berglund, { Stringy corrections to {K}aehler
  potentials, {SUSY} breaking, and the cosmological constant problem},  { JHEP}
  { 11} (2004) 085,
[\href{http://xxx.lanl.gov/abs/hep-th/0408054}{{\tt hep-th/0408054}}].

\bibitem{Berg:2005yu}
M.~Berg, M.~Haack, and B.~K{\"o}rs, { On volume stabilization by quantum
  corrections},  { Phys. Rev. Lett.} { 96} (2006) 021601,
[\href{http://xxx.lanl.gov/abs/hep-th/0508171}{{\tt hep-th/0508171}}].

\bibitem{Blumenhagen:2006ci}
R.~Blumenhagen, B.~K{\"o}rs, D.~L{\"u}st, and S.~Stieberger, { Four-dimensional
  string compactifications with {D}-branes, orientifolds and fluxes},
\href{http://xxx.lanl.gov/abs/hep-th/0610327}{{\tt hep-th/0610327}}.

\bibitem{Green:1999qt}
M.~B. Green, { Interconnections between type {II} superstrings, {M} theory and
  {N = 4} {Y}ang-{M}ills},
\href{http://xxx.lanl.gov/abs/hep-th/9903124}{{\tt hep-th/9903124}}.

\bibitem{Choi:2007ka}
K.~Choi and H.~P. Nilles, { The gaugino code},  { JHEP} { 04} (2007) 006,
[\href{http://xxx.lanl.gov/abs/hep-ph/0702146}{{\tt hep-ph/0702146}}].

\bibitem{vonGersdorff:2005bf}
G.~von Gersdorff and A.~Hebecker, { Kaehler corrections for the volume modulus
  of flux compactifications},  { Phys. Lett.} { B624} (2005) 270--274,
[\href{http://xxx.lanl.gov/abs/hep-th/0507131}{{\tt hep-th/0507131}}].

\bibitem{Nilles:1983ge}
H.~P. Nilles, { Supersymmetry, supergravity and particle physics},  { Phys.
  Rept.} { 110} (1984)
1.

\bibitem{Martin:1997ns}
S.~P. Martin, { A supersymmetry primer},
\href{http://xxx.lanl.gov/abs/hep-ph/9709356}{{\tt hep-ph/9709356}}.

\bibitem{Kaplunovsky:1987rp}
V.~S. Kaplunovsky, { One loop threshold effects in string unification},  {
  Nucl. Phys.} { B307} (1988) 145,
[\href{http://xxx.lanl.gov/abs/hep-th/9205068}{{\tt hep-th/9205068}}].

\bibitem{Conlon:2006tj}
J.~P. Conlon, D.~Cremades, and F.~Quevedo, { Kaehler potentials of chiral
  matter fields for {C}alabi-{Y}au string compactifications},  { JHEP} { 01}
  (2007) 022,
[\href{http://xxx.lanl.gov/abs/hep-th/0609180}{{\tt hep-th/0609180}}].

\bibitem{Berenstein:2006aj}
D.~Berenstein, { Branes vs. {GUTS}: Challenges for string inspired
  phenomenology},
\href{http://xxx.lanl.gov/abs/hep-th/0603103}{{\tt hep-th/0603103}}.

\bibitem{Ibanez:2001nd}
L.~E. Ibanez, F.~Marchesano, and R.~Rabadan, { Getting just the standard model
  at intersecting branes},  { JHEP} { 11} (2001) 002,
[\href{http://xxx.lanl.gov/abs/hep-th/0105155}{{\tt hep-th/0105155}}].

\bibitem{Candelas:1993dm}
P.~Candelas, X.~De~La~Ossa, A.~Font, S.~H. Katz, and D.~R. Morrison, { Mirror
  symmetry for two parameter models. {I}},  { Nucl. Phys.} { B416} (1994)
  481--538,
[\href{http://xxx.lanl.gov/abs/hep-th/9308083}{{\tt hep-th/9308083}}].

\bibitem{Green:1996dd}
M.~B. Green, J.~A. Harvey, and G.~W. Moore, { I-brane inflow and anomalous
  couplings on {D}-branes},  { Class. Quant. Grav.} { 14} (1997) 47--52,
[\href{http://xxx.lanl.gov/abs/hep-th/9605033}{{\tt hep-th/9605033}}].

\bibitem{Dasgupta:1997cd}
K.~Dasgupta, D.~P. Jatkar, and S.~Mukhi, { Gravitational couplings and {Z}(2)
  orientifolds},  { Nucl. Phys.} { B523} (1998) 465--484,
[\href{http://xxx.lanl.gov/abs/hep-th/9707224}{{\tt hep-th/9707224}}].

\bibitem{Cheung:1997az}
Y.-K.~E. Cheung and Z.~Yin, { Anomalies, branes, and currents},  { Nucl. Phys.}
  { B517} (1998) 69--91,
[\href{http://xxx.lanl.gov/abs/hep-th/9710206}{{\tt hep-th/9710206}}].

\bibitem{Minasian:1997mm}
R.~Minasian and G.~W. Moore, { K-theory and {R}amond-{R}amond charge},  { JHEP}
  { 11} (1997) 002,
[\href{http://xxx.lanl.gov/abs/hep-th/9710230}{{\tt hep-th/9710230}}].

\bibitem{Morales:1998ux}
J.~F. Morales, C.~A. Scrucca, and M.~Serone, { Anomalous couplings for
  {D}-branes and {O}-planes},  { Nucl. Phys.} { B552} (1999) 291--315,
[\href{http://xxx.lanl.gov/abs/hep-th/9812071}{{\tt hep-th/9812071}}].

\bibitem{Stefanski:1998yx}
J.~Stefanski, Bogdan, { Gravitational couplings of d-branes and o-planes},  {
  Nucl. Phys.} { B548} (1999) 275--290,
[\href{http://xxx.lanl.gov/abs/hep-th/9812088}{{\tt hep-th/9812088}}].

\bibitem{Bachas:1999um}
C.~P. Bachas, P.~Bain, and M.~B. Green, { Curvature terms in {D}-brane actions
  and their {M}-theory origin},  { JHEP} { 05} (1999) 011,
[\href{http://xxx.lanl.gov/abs/hep-th/9903210}{{\tt hep-th/9903210}}].

\bibitem{Fotopoulos:2001pt}
A.~Fotopoulos, { On (alpha')**2 corrections to the {D}-brane action for non-
  geodesic world-volume embeddings},  { JHEP} { 09} (2001) 005,
[\href{http://xxx.lanl.gov/abs/hep-th/0104146}{{\tt hep-th/0104146}}].

\bibitem{Dudas:2004nd}
E.~Dudas, G.~Pradisi, M.~Nicolosi, and A.~Sagnotti, { On tadpoles and vacuum
  redefinitions in string theory},  { Nucl. Phys.} { B708} (2005) 3--44,
[\href{http://xxx.lanl.gov/abs/hep-th/0410101}{{\tt hep-th/0410101}}].

\bibitem{Kaloper:1999yr}
N.~Kaloper and R.~C. Myers, { The {O}(dd) story of massive supergravity},  {
  JHEP} { 05} (1999) 010,
[\href{http://xxx.lanl.gov/abs/hep-th/9901045}{{\tt hep-th/9901045}}].

\bibitem{Kachru:2002he}
S.~Kachru, M.~B. Schulz, and S.~Trivedi, { Moduli stabilization from fluxes in
  a simple {IIB} orientifold},  { JHEP} { 10} (2003) 007,
[\href{http://xxx.lanl.gov/abs/hep-th/0201028}{{\tt hep-th/0201028}}].

\bibitem{Berg:2004ek}
M.~Berg, M.~Haack, and B.~K{\"o}rs, { Loop corrections to volume moduli and
  inflation in string theory},  { Phys. Rev.} { D71} (2005) 026005,
[\href{http://xxx.lanl.gov/abs/hep-th/0404087}{{\tt hep-th/0404087}}].

\bibitem{Berg:2004sj}
M.~Berg, M.~Haack, and B.~K{\"o}rs, { On the moduli dependence of
  nonperturbative superpotentials in brane inflation},
\href{http://xxx.lanl.gov/abs/hep-th/0409282}{{\tt hep-th/0409282}}.

\bibitem{Lust:2004cx}
D.~L{\"u}st, P.~Mayr, R.~Richter, and S.~Stieberger, { Scattering of gauge,
  matter, and moduli fields from intersecting branes},  { Nucl. Phys.} { B696}
  (2004) 205--250,
[\href{http://xxx.lanl.gov/abs/hep-th/0404134}{{\tt hep-th/0404134}}].

\bibitem{Burgess:1986ah}
C.~P. Burgess and T.~R. Morris, { Open and unoriented strings a la {P}olyakov},
   { Nucl. Phys.} { B291} (1987)
256.

\bibitem{Burgess:1986wt}
C.~P. Burgess and T.~R. Morris, { Open superstrings a la {P}olyakov},  { Nucl.
  Phys.} { B291} (1987)
285.

\end{thebibliography}\endgroup

\end{document}